\documentclass[a4paper,11pt]{article}
\usepackage{jheppub} % for details on the use of the package, please see the JINST-author-manual
\usepackage{lineno}
\usepackage{enumitem}
\title{\boldmath Mixing ``Magnetic'' and ``Electric'' Ehlers--Harrison transformations: The Electromagnetic Swirling Spacetime and Novel Type I Backgrounds}

\author[a]{Jos\'e Barrientos,}
\author[a,b]{Adolfo Cisterna,}
\author[b]{Ivan Kol\'a{\v r},}
\author[c]{Keanu M\"uller,}
\author[c]{Marcelo Oyarzo,}
\author[d]{and Konstantinos Pallikaris}
 
 \affiliation[a]{Sede Esmeralda, Universidad de Tarapac\'a, Avenida Luis Emilio Recabarren 2477, Iquique, Chile}
 \affiliation[b]{Institute of Theoretical Physics, Faculty of Mathematics and Physics, Charles University, V Hole{\v s}ovi{\v c}k{\'a}ch 2, 180 00 Prague 8, Czech Republic}
  \affiliation[c]{Departamento de F\'isica, Universidad de Concepci\'on, Casilla, 160-C, Concepci\'on, Chile}
 \affiliation[d]{Laboratory of Theoretical Physics, Institute of Physics, University of Tartu, W. Ostwaldi 1, 50411 Tartu, Estonia}

\emailAdd{jbarrientos@academicos.uta.cl, adolfo.cisterna@mff.cuni.cz,\\ ivan.kolar@matfyz.cuni.cz, keanumuller2016@udec.cl,\\moyarzo2016@udec.cl, konstantinos.pallikaris@ut.ee}

\abstract{In this paper, we obtain a complete list of stationary and axisymmetric spacetimes, generated from a Minkowski spacetime using the Ernst technique. We do so by operating on the associated seed potentials with a composition of Ehlers and Harrison transformations. In particular, assigning an additional ``electric'' or ``magnetic'' tag to the transformations, we investigate the new spacetimes obtained either via a composition of magnetic Ehlers and Harrison transformations (first part) or via a magnetic-electric combination (second part). In the first part, the resulting type D spacetime, dubbed electromagnetic swirling universe, features key properties, separately found in swirling and (Bonnor--)Melvin spacetimes, the latter recovered in appropriate limits. A detailed analysis of the geometry is included, and subtle issues are addressed. A detailed proof that the spacetime belongs to the Kundt family, is included, and a notable relation to the planar-Reissner-Nordstr\"om-NUT black hole is also meticulously worked out. This relation is further exploited to reverse-engineer the form of the solution in the presence of a nontrivial cosmological constant. A Schwarzschild black hole embedded into the new background is also discussed. In the second part, we present four novel stationary and axisymmetric asymptotically nonflat type I spacetimes, which are naively expected to be extensions of the Melvin or swirling solution including a NUT parameter or electromagnetic charges. We actually find that they are, under conditions, free of curvature and topological singularities, with the physical meaning of the electric transformation parameters in these backgrounds requiring further investigation.}

\renewcommand{\Re}{\operatorname{Re}}
\renewcommand{\Im}{\operatorname{Im}}

\newcommand{\bs}[1]{\boldsymbol{#1}}
\newcommand{\ernst}{\mathcal{E}}

\begin{document}
\maketitle
\flushbottom
\section{Introduction}
Since the advent of Einstein's equations, the quest for exact solutions to this set of coupled nonlinear partial differential equations has played an important part in modern physics \cite{Stephani:2003tm}. Exact solutions have significantly aided us in understanding many classical and semiclassical properties of gravity, and this is exactly what makes them integral to our comprehension of diverse phenomena occurring at astrophysical and cosmological scales. Furthermore, these solutions have set the stage for the theoretical exploration of many groundbreaking concepts, including, but not limited to, black hole thermodynamics \cite{Bekenstein:1972tm,Bekenstein:1973ur,Hawking:1974rv,Hawking:1975vcx}, the information paradox \cite{Giddings:1995gd}, and holography \cite{Susskind:1994vu}.

Dealing with the field equations of General Relativity (GR) poses a nontrivial challenge, which reduces to a tractable problem only when a high amount of symmetry is imposed. In particular, exploiting the Lie point symmetries inherent in Einstein's equations, serves as a robust method for generating exact solutions---solutions that would be practically impossible to integrate with brute force.
Two especially interesting Lie point symmetries of the Einstein--Maxwell system, Ehlers~\cite{Ehlers:1957zz,Ehlers:1959aug} and Harrison~\cite{harrison1968new} transformations, suffice for the construction of novel stationary and axially symmetric spacetimes. These two transformations are part of a larger set of Lie point symmetries, which exist as hidden symmetries of the Einstein--Maxwell system of field equations, and which are revealed only when one formulates the theory in terms of the complex so-called Ernst potentials \cite{Ernst:1967wx,Ernst:1967by}; they are indeed potential-space symmetries, the parameters of which comprise an eight-parameter isometry group, a nonlinear representation of which was originally given in~\cite{NeugeKramer69}. The linear representation of this group and the apparent isomorphism with SU(2,1), was only a few years later delivered by Kinnersley~\cite{Kinn1}, who extended the previous results of Geroch~\cite{Geroch} to include electromagnetism. 

Recently, stationary and axisymmetric spacetimes have received a considerable amount of attention within the framework of the Ernst description~\cite{Ernst:1967wx,Ernst:1967by} of gravity.\footnote{These equations were originally derived by Ernst in~\cite{Ernst:1967by} for stationary electrovac fields with the further assumption of axisymmetry. Later, Israel and Wilson~\cite{Israel:1972vx}, as well as Harrison~\cite{harrison1968new}, rederived them independently for general stationary electrovac fields.} In particular, the significant effect that Ehlers and Harrison transformations have on accelerating spacetimes, has been explored in detail~\cite{Barrientos:2023tqb,Astorino:2023elf,Astorino:2023ifg,Barrientos:2023dlf}. Using Ehlers or Harrison transformations, or a combination of them for that matter, it has lately been demonstrated that certain algebraically \textit{special} accelerating spacetimes can be mapped to novel algebraically \textit{general} solutions. For example, some of us have successfully constructed a complete hierarchy of type I spacetimes~\cite{Barrientos:2023dlf} obtained via this generating technique, generalizing the well-known Pleba\'nski--Demia\'nski family, i.e., the most general family of type D solutions in Einstein--Maxwell theory. 

To better understand these developments, we shall briefly review the effects of these transformations. For seed spacetimes cast into the electric form of the Weyl--Lewis--Papapetrou (WLP) metric,\footnote{See the next section for definitions of ``electric'' and ``magnetic'' forms.} the standard lore is that Ehlers transformations introduce an additional real parameter which in certain cases can be associated with a NUT parameter in the target metric. On the other hand, Harrison transformations introduce an additional complex parameter, whose real and imaginary parts, can in certain cases be associated with monopolic electromagnetic charges in the target configuration. When the seed is a type D accelerating spacetime, it has been observed that, on top of the above effects, the transformations also bring about a change in the algebraic character of the generated solution, namely the target spacetimes are of type I. An explanation for this peculiar effect seemingly lies in the way that the transformation parameters enter the new metrics. 

To get a good grasp on this, let us for a moment consider the Schwarzschild black hole as a prototypical seed example. Casting it into the electric form of the WLP metric and operating on the seed potentials with an Ehlers transformation, it is known that the resulting spacetime is the Taub--NUT black hole, modulo coordinate transformations, and parameter redefinitions. In other words, the transformation---without changing the Petrov type~\cite{Barrientos:2023tqb,Astorino:2023elf}---has introduced a new parameter, which now, together with the mass, determines the location of the black hole horizon. In sharp contrast to this, if we let our seed be the $C$-metric~\cite{newman1961new,robinson1962some,witten1962gravitation}, an Ehlers transformation will not only affect the usual event horizon in the above sense, but also the Rindler one~\cite{Barrientos:2023dlf,Astorino:2023ifg}. As a byproduct of this, the resulting spacetime turns out to be an accelerating Schwarzschild black hole with a NUT-like parameter, which however is of type I, and as such, it cannot be found within the Pleba\'nski--Demia\'nski type D hierarchy. Similarly, a Harrison transformation has its two real parameters entering both, black hole and Rindler, horizons of the $C$-metric, thereby leading to a charged accelerating black hole of type I and not in a Reissner--Nordstr\"om-$C$-metric of type D, as one may have perhaps expected. 

Important insight into the modified Rindler horizons in these type I accelerating black holes can be obtained by viewing the new solutions as a particular limit of black hole binaries\cite{Astorino:2023ifg}. Recall that the near-horizon geometry of a Schwarzschild black hole is described by the Rindler metric, characteristic of an accelerating observer. A mathematically equivalent way to ``zoom in'' on the event horizon is to take the infinite mass limit of the solution. An accelerating Schwarzschild black hole can be conceptualized as a binary system of two Schwarzschild black holes, effectively described by the Bach--Weyl solution \cite{BachWeyl}, where one of the two grows infinitely large (becomes infinitely massive) while retaining a finite distance from the other. The event horizon of the ``big'' black hole appears then as an accelerating horizon to its small sibling. Consequently, the Bach-Weyl solution ends up appearing as a $C$-metric in this limit. Analogously, these type I accelerating black holes, featuring Rindler horizons which depend on the transformation parameters among others, can be thought about as a limit of the NUTty and/or charged extension of the Bach-Weyl spacetime. A complete hierarchy of these novel type I spacetimes, including also a seed angular momentum parameter, can be found in~\cite{Barrientos:2023dlf}.\footnote{The solution with angular momentum has also later appeared in the fifth revision of~\cite{Astorino:2023ifg}.}

In light of these recent developments, this study aims to shed light on the remaining spacetimes one can generate by composing Ehlers and Harrison transformations. To further elaborate on our agenda, it is best if we use the following terminology, which will be formally introduced in Sec.~\ref{sec:ErnstForma}. An \textit{electric} (resp. \textit{magnetic}) Ehlers transformation is an Ehlers transformation of the Ernst potentials associated with a seed metric cast into the electric (resp. magnetic) form of the WLP metric, and ditto for Harrison. To date, and to the best of our knowledge, only the combination of electric transformations has been investigated. In this work, we wish to fill the gaps, and we do so by first combining a magnetic Ehlers transformation with a magnetic Harrison one (first part), and then by taking all possible combinations of an electric and a magnetic transformation (second part). It is a firmly established fact that operating with magnetic Harrison and magnetic Ehlers transformations on the seed potentials of Minkowski spacetime, one obtains two interesting asymptotically nontrivial backgrounds, commonly known as Bonnor--Melvin~\cite{Bonnor_1954,Melvin1966} and swirling~\cite{Astorino:2022aam, Gibbons:2013yq} spacetimes, respectively. We will briefly review them later on. In the first part of this work, we present a more general spacetime, dubbed electromagnetic swirling universe, from which, as the name suggests, the aforementioned solutions follow in appropriate limits. We study this geometry in detail, also addressing a rather subtle issue concerning the uniqueness of timelike Killing vectors, relevant also in the swirling case, which has been unfortunately neglected so far. Moreover, we prove that the new metric is Kundt via an intricate chain of coordinate reparametrizations and parameter redefinitions. We also explicitly prove the existence of an intriguing relation between this new background and a planar Reissner--Nordstr\"om--NUT spacetime. It is this particular relation which we exploit to also analytically derive the cosmological extension of the electromagnetic swirling universe, i.e., the form of the solution in the presence of a cosmological constant. As a finale to the first part, we embed a Schwarzschild black hole into the new background, giving some emphasis on the dragging effect and the deformation of the horizon surface. 

In the second part, we give attention to electric-magnetic mixtures, registering the complete list of novel spacetimes one can generate from Minkowski spacetime by combining an electric with a magnetic transformation. Four type I families are obtainable in this way, and we discuss the conditions under which they can be legitimately called backgrounds. Due to the number of solutions, the analysis will not be exhaustive. We show that all spacetimes in the second part of this work, feature closed timelike/null curves (CTCs/CNCs). These appear inside regions, the boundaries of which are also surfaces where the frame-dragging angular velocity becomes singular. We then argue that one may perhaps ascribe the occurrence of nonchronal regions to the intensity of rotation building up very close to these singular surfaces; inertial frames get dragged so strongly that the light cones end up being tilted in the direction of the circumference. We remark that two of the backgrounds carry everywhere finite electric and magnetic fields which decay in all directions at infinity. Such behavior is in stark contrast to what happens in the Bonnor--Melvin solution, where the fields are uniform in the vicinity of the symmetry axis. Whether CTC-free parts of these two backgrounds could perhaps be as suitable as the former spacetime, for example in describing astrophysical black holes surrounded by strong magnetic fields, remains an open question. The presence of closed timelike/null curves requires much deeper scrutiny. Although they render the entirety of each solution unrealistic for modeling (astro)physical phenomena, the fact that these pathologies beset solutions to the Einstein--Maxwell theory, perhaps makes these backgrounds interesting from a totally different perspective. 

This paper is structured as follows. In Sec.~\ref{sec:ErnstForma} we communicate the very basics of the Ernst formalism and briefly discuss the symmetry transformations. We use this section to also derive the nomenclature, and we further lay down the steps we follow to generate new solutions, giving an algorithmic description of the generating technique we use in this work. In Sec.~\ref{sec:Petrov}, we review, in double-quick time, a highly convenient method for the Petrov classification, which we will use throughout. In Sec.~\ref{sec:MS}, we combine magnetic Ehlers and Harrison transformations to obtain the electromagnetic swirling universe, analyzing its geometric properties and making its relation to a planar Reissner-Nordstr\"om-NUT spacetime manifest. We use the latter link to derive the electromagnetic-swirling-(A)dS solution. Finally, a Schwarzschild black hole is embedded into the new background, with its most interesting features discussed in some depth. In Sec.~\ref{mixing}, we direct our efforts towards generating new spacetimes by combining electric and magnetic Ehlers and Harrison transformations. We start with a Minkowski seed and work our way up to the four different spacetimes one can obtain by combining a magnetic transformation with an electric one. We investigate whether (and under which conditions) these asymptotically nonflat geometries can be characterized as backgrounds. Finally, in Sec.~\ref{sec:Conclusions} we summarize our findings and conclude, also suggesting new possible avenues for further research.

\section{The essentials}
\subsection{The complex-potential formalism\label{sec:ErnstForma}}
In this section, we review the formulation of the Einstein--Maxwell field equations in terms of two complex potentials, as presented by Ernst in his seminal works~\cite{Ernst:1967wx,Ernst:1967by}. Making a stationary and axisymmetric ansatz, one can introduce two complex potentials, $\ernst$, and $\Phi$, and write the Einstein--Maxwell field equations as a pair of complex equations. Doing so, a set of Lie point symmetries is revealed, the realization of which eludes one in the usual tensorial formalism. This set of symmetry transformations can then be exploited to generate new solutions from old ones. Let us briefly review the scheme. 

The first step is to make a stationary and axisymmetric metric ansatz, the Weyl--Lewis--Papapetrou (WLP) ansatz,
\begin{equation}
    ds^2 = -f(dt-\omega\,d\phi)^2+f^{-1}[\mathrm{e}^{2\gamma}(d\rho^2+dz^2)+\rho^2 d\phi^2],\label{eq:eleWLP}
\end{equation}
where $f,\ \omega,$ and $\gamma$ are functions of $\rho$ and $z$. We further take our gauge field to have the symmetry-compatible form 
\begin{equation}
    A=A_t\,dt+A_\phi\,d\phi,\label{eq:GFansatz}
\end{equation}
where the scalar potentials are also functions of the Weyl coordinates $\rho$ and $z$. Defining 
\begin{equation}
    \ernst := f-\Phi\overline{\Phi}+i\chi,\quad \Phi:=A_t+i\widetilde{A}_\phi,\label{eq:ElePots}
\end{equation}
one may show, after some cumbersome algebra, that the Einstein--Maxwell field equations for stationary axisymmetric fields are equivalent to the complex Ernst equations 
\begin{subequations}
\label{eq:ernsteqs}
\begin{eqnarray}
(\Re\ernst+\Phi\overline{\Phi})\nabla^2 \mathcal{E}&=&\bs{\nabla} \ernst\cdot (\bs{\nabla} \ernst +2 \overline{\Phi} \bs{\nabla} \Phi), \\
(\Re\ernst+\Phi\overline{\Phi}) \nabla^2 \Phi&=&\bs{\nabla} \Phi \cdot (\bs{\nabla}\ernst+2 \overline{\Phi} \bs{\nabla} \Phi),
\end{eqnarray}
\end{subequations}
together with a pair of equations determining $\gamma$ via integration by quadrature.\footnote{These will not bother us, for the symmetry transformations do not transform $\gamma$ at all.} The Laplacian and the gradient are understood as operators in three-dimensional Euclidean space in cylindrical coordinates. The potentials $\chi$ and $\widetilde{A}_\phi$ are twist potentials satisfying the real equations
\begin{subequations}
\begin{eqnarray}
\hat{\phi} \times \bs{\nabla} \chi&=&-\rho^{-1} f^2 \bs{\nabla} \omega-2 \hat{\phi} \times \Im(\overline{\Phi} \bs{\nabla} \Phi), \label{eq:eqchiE}\\
\hat{\phi} \times \bs{\nabla} \widetilde{A}_{\phi}&=&\rho^{-1} f(\bs{\nabla} A_{\phi}+\omega {\bs{\nabla}A_t}),\label{eq:eqATphiE}
\end{eqnarray}
\end{subequations}
respectively, where $\hat{\phi}$ is the unit normal in the azimuthal direction. Interestingly, if we define our complex potentials as 
\begin{equation}
    \ernst := -f-\Phi\overline{\Phi}-i\chi,\quad \Phi:=A_\phi +i \widetilde{A}_t,\label{eq:MagPots}
\end{equation}
and considering the metric ansatz 
\begin{equation}
    ds^2 = f(d\phi-\omega\,dt)^2+f^{-1}[\mathrm{e}^{2\gamma}(d\rho^2+dz^2)-\rho^2 dt^2],\label{eq:mWLP}
\end{equation}
instead of~\eqref{eq:eleWLP}, then for a gauge field as in eq.~\eqref{eq:GFansatz}, we also arrive at eqs.~\eqref{eq:ernsteqs}, with the twist potentials now satisfying the equations 
\begin{subequations}
\begin{eqnarray}
\hat{\phi} \times \bs{\nabla} \chi&=&-\rho^{-1} f^2 \bs{\nabla} \omega+2 \hat{\phi} \times \Im(\overline{\Phi} \bs{\nabla} \Phi), \label{eq:eqchiM}\\
\hat{\phi} \times \bs{\nabla} \widetilde{A}_{t}&=&\rho^{-1} f(\bs{\nabla} A_{t}+\omega {\bs{\nabla}A_\phi}).\label{eq:eqATt}
\end{eqnarray}
\end{subequations}

As previously mentioned, eqs.~\eqref{eq:ernsteqs} are invariant under a bunch of symmetry transformations in potential space, whose domain and target (as maps) are potentials associated with stationary and axisymmetric Einstein--Maxwell fields. Their finite forms read~\cite{NeugeKramer69}
\begin{subequations}\label{eq:Symmetries}
    \begin{eqnarray}
        \operatorname{G_1}[a]:(\ernst_0,\Phi_0)\mapsto (\ernst,\Phi)&:=&(\ernst + i a,\Phi_0),\\
        \operatorname{G_2}[\alpha]:(\ernst_0,\Phi_0)\mapsto (\ernst,\Phi)&:=&(\ernst_0 - 2\overline{\alpha} \Phi_0 - \alpha\overline{\alpha},\Phi_0+\alpha),\\
        \operatorname{D}[\epsilon]:(\ernst_0,\Phi_0)\mapsto (\ernst,\Phi)&:=&(\epsilon\overline{\epsilon}\ernst_0,\epsilon\Phi_0),\\
        \operatorname{E}[c]:(\ernst_0,\Phi_0)\mapsto (\ernst,\Phi)&:=&\frac{(\ernst_0,\Phi_0)}{1+ic\ernst_0},\\
        \operatorname{H}[\beta]:(\ernst_0,\Phi_0)\mapsto (\ernst,\Phi)&:=&\frac{(\ernst_0,\Phi_0+\beta\ernst_0)}{1-2\overline{\beta}\Phi_0-\beta\overline{\beta}\ernst_0},
    \end{eqnarray}
\end{subequations}
where $a,c$ are real parameters and $\alpha,\beta,\epsilon$ complex. Transformations $\operatorname{G}_1$ and $\operatorname{G}_2$ are ``gauge'' transformations which transform the potentials, but leave the metric and the gauge field invariant, $\operatorname{D}$ is a duality-rescaling transformation, $\operatorname{E}$ denotes the Ehlers transformation~\cite{Ehlers:1957zz}, and $\operatorname{H}$ stands for the Harrison~\cite{harrison1968new} one. Note that a composition of $\operatorname{G}_1$, $\operatorname{D}$ and $\operatorname{E}$ gives the inversion map
\begin{equation}
    \operatorname{I}:(\ernst_0,\Phi_0)\mapsto (\ernst,\Phi):=\frac{(1,\Phi_0)}{\ernst_0},
\end{equation}
which thus is also a (discrete) symmetry of the Ernst equations. In particular, 
\begin{equation}
    \operatorname{I}=\operatorname{G}_1[a]\circ\operatorname{E}[1/a]\circ\operatorname{G}_1[a]\circ\operatorname{D}[ia],
\end{equation}
and we can easily verify that 
\begin{equation}
    \operatorname{E}[a]=\operatorname{I}\circ \operatorname{G}_1[a]\circ\operatorname{I},\quad \operatorname{H}[\alpha]=\operatorname{I}\circ \operatorname{G}_2[\alpha]\circ\operatorname{I}.
\end{equation}

The transformations~\eqref{eq:Symmetries} are associated with eight Killing vectors (KVs) locally generating an isometry group of the potential space whose linear representation is a representation of SU(2,1)~\cite{Kinn1}. As discussed in~\cite{Barrientos:2023dlf}, Ehlers transformations form a one-dimensional subgroup, i.e., $\operatorname{E}[a]\circ\operatorname{E}[b]=\operatorname{E}[a+b]$, and they commute with Harrison transformations because $[\operatorname{G}_1,\operatorname{G}_2]=0$. On the other hand, two Harrison transformations do not in general yield a Harrison, simply because $[\operatorname{G}_2,\operatorname{G}_2]\neq 0$. Actually, 
\begin{equation}
    \operatorname{H}[\alpha]\circ \operatorname{H}[\beta] = \operatorname{E}[-2\Im(\alpha\overline{\beta})]\circ\operatorname{H}[\alpha+\beta],\label{eq:HHtoEH}
\end{equation}
so, unless $\alpha\overline{\beta}=\beta\overline{\alpha}$, a ``Harrison of Harrison'' gives an ``Ehlers of Harrison'' with the Ehlers parameter fixed in terms of the Harrison parameters $\alpha$ and $\beta$. If we recall that the Harrison map generates electrovac solutions from vacuum ones, then eq.~\eqref{eq:HHtoEH} implies that two Harrison transformations map static vacuum solutions to \textit{stationary} electrovac ones. 

Let us also agree on the terminology to be used in this work. The metric~\eqref{eq:eleWLP} will be called the electric form of the WLP, whereas~\eqref{eq:mWLP} will be the magnetic form. These names~\cite{Vigano:2022hrg}, which unfortunately are misleading, are based on the simple observation that a Harrison transformation (with real parameter) of the potentials~\eqref{eq:ElePots} associated with a vacuum spacetime cast into~\eqref{eq:eleWLP}, introduces an electric charge, whereas the same transformation acting on the potentials~\eqref{eq:MagPots} associated with the very same vacuum spacetime cast into~\eqref{eq:mWLP}, gives new potentials associated with a magnetized version of the seed. Nevertheless, this way of naming things is convenient for the task at hand, and we adhere to it henceforth. Therefore, the potentials~\eqref{eq:ElePots} will be called electric, as they are associated with a seed metric cast into the electric WLP, while the potentials~\eqref{eq:MagPots} will be dubbed magnetic by the same reasoning. Finally, a symmetry transformation of electric potentials will be addressed as electric, and ditto for the magnetic case. 

Since this is a solution-generating technique, the process is purely algorithmic. To the aid of the interested reader, we list the steps below for the so-called electric case. The same steps ought to be followed in the magnetic case using the relevant equations.  
\begin{enumerate}
    \item Given a stationary and axisymmetric metric, identify $f,\ \omega,$ and $\gamma$ in~\eqref{eq:eleWLP} via direct comparison.
    \item These should be fed to eq.~\eqref{eq:eqATphiE}, together with the components of the seed gauge field, which can then be integrated for the twist potential $\widetilde{A}_\phi$. Results should in turn be fed to eq.~\eqref{eq:eqchiE} to obtain $\chi$. 
    \item Substitute everything into the definitions~\eqref{eq:ElePots} to get exact expressions for the seed potentials. Next, do your favorite transformations and read off the target functions $f,\ A_t,\ \Tilde{A}_\phi,$ and $\chi$ again from~\eqref{eq:ElePots}.
    \item These should now be fed to eq.~\eqref{eq:eqchiE}, which can be integrated for $\omega$. Having $\omega$, plug everything into eq.~\eqref{eq:eqATphiE} to get the azimuthal component of the target gauge field.
    \item Substitute the target $f,\ \omega,\ A_t,\ A_\phi,$ and the seed $\gamma$ into the metric~\eqref{eq:eleWLP} and the gauge field~\eqref{eq:GFansatz}, and voil\`a. 
\end{enumerate}
Although this procedure looks simple per se, the computational complexity involved may become quite intractable. 

\subsection{Petrov classification\label{sec:Petrov}}
A fundamental way to distinguish gravitational fields, independent of the coordinate system, is to classify them according to their Petrov type~\cite{Petrov2000TheCO}. To do so, we need to study the algebraic structure of the Weyl tensor, which in four dimensions has ten independent components, encoded in the five complex Weyl--NP scalars $\Psi_0,\ldots,\Psi_4$, within the framework of the Newman--Penrose formalism. The Petrov classification is a useful tool in our case, for one can directly prove the possible nonequivalence between certain solutions appearing in this manuscript, by simply looking at their Petrov type. This being the case, it is worth including a double-quick review of the classification algorithm used in this work.

First, we consider an arbitrary complex null tetrad (CNT) $\mathfrak{e}=\{k,l,m,\overline{m}\}$ with $k,l$ real null vectors and $m,\overline{m}$ complex conjugate null vectors, such that $k^a l_a=-1$, $m^a \overline{m}_a=1$, and all other products zero. Latin indices are lowered/raised with the use of the metric $g_{ab}=g_{\mu\nu}\mathfrak{e}_a^\mu\mathfrak{e}_b^\nu$ and its inverse. Having a complex null tetrad, we next consider the definition of the five complex Weyl--NP scalar as given in~\cite{Stephani:2003tm}. The problem of finding the Petrov type of a given spacetime will be attacked using the d'Inverno and Russell-Clark method~\cite{dInverno}. Starting with an arbitrary null basis, we wish to find the specific Lorentz transformation leading to a new basis, in which the number of vanishing Weyl scalars is maximal. For $\Psi_4\neq 0$, it is known that this is equivalent to the problem of finding the roots of the complex quartic equation
\begin{equation}
    \Psi_4 \epsilon^4 - 4\Psi_3\epsilon^3 + 6\Psi_2 \epsilon^2-4\Psi_1\epsilon+\Psi_0=0.\label{eq:quartic}
\end{equation}

Based on the number and multiplicity of these roots, we can then determine the Petrov type. To make things simpler, we will choose our CNT such that $\Psi_1=0=\Psi_3$ and $\Psi_4\neq 0$. In App.~\ref{app:ON&CNtet}, we show that such a choice is always possible for the general WLP metric, and we explicitly suggest the way to construct it. Given this CNT, eq.~\eqref{eq:quartic} becomes a quadratic for $z=\epsilon^2$, and dividing it by $\Psi_4$, we get 
\begin{equation}
    z^2+\frac{6\Psi_2}{\Psi_4}z+\frac{\Psi_0}{\Psi_4}=0,\label{eq:quadratic}
\end{equation}
with discriminant proportional to
\begin{equation}
    9\Psi_2^2-\Psi_0\Psi_4.
\end{equation}
The two roots of the quadratic are 
\begin{equation}
    z_{\pm}=\frac{-3\Psi_2\pm \sqrt{9\Psi_2^2-\Psi_0\Psi_4}}{\Psi_4}.
\end{equation}
Thus, if $9\Psi_2^2 \neq \Psi_0\Psi_4$, the quartic eq.~\eqref{eq:quartic} has four simple roots $\{\pm\sqrt{z_+},\pm\sqrt{z_-}\}$, meaning that the Petrov type is I. On the other hand, if the discriminant vanishes, the quadratic has a double root $z_0=-3\Psi_2/\Psi_4$ which implies that the quartic eq.~\eqref{eq:quartic} has two double roots $\pm \sqrt{z_0}$. This means that the Petrov type is D in this case. Clearly, all spacetimes in the WLP family will be either O, D, or I. Having completed the formalities, we shall now proceed with the construction and study of a Schwarzschild black hole embedded into an electromagnetic swirling universe, with enough emphasis given also on the underlying background geometry.  

\section{The electromagnetic swirling universe}
\label{sec:MS}

Since this section is dedicated to a sort of composite background, we shall start it with a quick discussion about the separate building blocks of the latter, the electromagnetic universe and the swirling spacetime. The magnetic universe, also known as the Bonnor--Melvin solution, was first found by Bonnor~\cite{Bonnor_1954}, and it was only later rediscovered by Melvin~\cite{Melvin1966}. It describes a static and cylindrically symmetric magnetic field immersed in its own gravitational field. In other words, it can be seen as describing a magnetic flux tube held together by its own gravitational pull. The magnetic field lines are parallel to the axis of symmetry, and the field can be treated as a uniform one only near the vicinity of the axis. Since it contributes to the stress tensor, and since stress-energy acts as a gravitating mass, its intensity must be falling off far away from the symmetry axis to prevent a collapse under its own gravity; and this is the case indeed. 

Here, we shall present the solution with both, magnetic and electric, external fields. We will refer to it as the electromagnetic universe. In cylindrical coordinates, the metric describing it reads
\begin{equation}
    ds^2_{\mathsf{EM}}=\frac{\rho^2}{V^2}d\phi^2+V^2(-dt^2+d\rho^2+dz^2),\label{eq:MelvinMet}
\end{equation}
where $V(\rho):=1+X\overline{X}\rho^2$ and $\overline{X}$ is the complex conjugate of $X:=(E+iB)/2$, with $E$ (resp. $B$) controlling the intensity of the electric (resp. magnetic) field. This metric is accompanied by the gauge field 
\begin{equation}
    A=-z E\,dt-\frac{B\rho^2}{2V}d\phi,\label{eq:MelvinGF}
\end{equation}
and it belongs to the Kundt class of Petrov type D electrovac spacetimes. Asymptotically, it approached the Levi-Civita spacetime\footnote{This is the Levi-Civita metric with $\sigma=1=k$. See~\cite{griffiths_podolsky_2009} for more details.}
\begin{equation}
    ds^2=\rho^{-2}d\phi^2 +\rho^4(-dt^2+d\rho^2+dz^2).
\end{equation}
To see this, a rescaling of the noncompact coordinates is necessary.

Its motion group is locally generated by the four KVs
\begin{equation}
  T_1=\partial_t,\quad  T_2=\partial_z,\quad  T_3=\partial_\phi, \quad T_4=z \partial_t+t \partial_z.
\end{equation}
These KVs do not commute, thus forming a nonabelian Lie algebra $\mathfrak{g}$ with nonvanishing brackets 
\begin{equation}
    [T_1,T_4]=T_2,\quad [T_2,T_4]=T_1.
\end{equation}
The center of this Lie algebra is the one-dimensional subspace $\operatorname{span}\{T_3\}$ which is obviously isomorphic to the real line. This is a solvable algebra with its derived subalgebra being two-dimensional abelian. One can actually observe that $\mathfrak{g}$ is a trivial extension of $\mathfrak{e}(1,1)$, the latter being the Lie algebra of the pseudo-Euclidean group $E(1,1)$ of rigid motions in Minkowski 2-space. 

Another interesting feature of this solution is that, much like what happens in anti-de Sitter space (AdS), timelike geodesics are forbidden to escape to radial infinity due to the strong attraction towards the axis of symmetry. Yet, here the source of this extreme gravitational pull is not a negative cosmological constant, but rather the electromagnetic field itself (see~\cite{Melvin1966} for the study of geodesic motion in the magnetic universe). Finally, let us remark that the electromagnetic universe can be obtained from a Minkowski seed via a magnetic Harrison transformation with parameter $i \overline{X}$. 

Definitely, less has been said about the swirling spacetime. This is a stationary vacuum solution of Einstein's field equations, 
\begin{eqnarray}\label{eq:swirlingMet}
    ds^2_{\mathsf{S}}=\frac{\rho^2}{1+j^2\rho^4}(d\phi+4j z\,dt)^2+(1+j^2\rho^4)(-dt^2+d\rho^2+dz^2),
\end{eqnarray}
with the above expression first reported in~\cite{Gibbons:2013yq}, to the best of our knowledge, as an analytic continuation of the Bianchi II cosmological metric of Taub~\cite{Taub:1950ez}. This metric belongs to the Kundt family of Petrov type D vacuum solutions,\footnote{Since type D vacuum solutions in Kundt's class were classified by Kinnersley in~\cite{Kinnersley1969TYPEDV}, this metric probably appears therein, though definitely in a different chart.} and its associated isometry group is locally generated by the four KVs
\begin{equation}\label{eq:swirlingKVs}
  T_1=\partial_t,\quad T_2=\partial_\phi, \quad T_3 =z \partial_t+t \partial_z-2j(t^2+z^2)\partial_\phi, \quad  T_4=\partial_z-4j t\partial_\phi.
\end{equation}
Besides them, there is an additional irreducible rank-2 Killing tensor obtained from the Killing--Yano 2-form 
\begin{equation}\label{eq:KYswirling}
    -4 \jmath \rho z d t \wedge d \rho+\jmath \rho^2\left(1+\jmath^2 \rho^4\right) d t \wedge d z+\rho d \rho \wedge d \varphi.
\end{equation}

Since the KVs do not commute, their linear span ought to be a nonabelian Lie algebra. To identify this algebra, it is best if we choose another set of basis vectors, $\{T_+,T_-,T,T_3\}$, with $T_\pm = ({T_1\pm T_4})/\sqrt{2}$ and $T=4j T_2$. Concerning the new basis, the nonvanishing Lie brackets read
\begin{equation}
     [T_\pm,T_3]=\pm T_\pm,\quad [T_+,T_-]=T,
\end{equation}
and it is now easy to see that the derived subalgebra, spanned by $\{T_{\pm},T\}$, is the three-dimensional Heisenberg algebra. It turns out that the full Lie algebra, with center $\operatorname{span}\{T\}$, is solvable and nondecomposable. If it bears a special name, then this name unfortunately eludes us. It features as ${A}_{4,8}$ in the classification of four-dimensional Lie algebras by Patera and Winternitz~\cite{PateraWinternitz}.

The limit of~\eqref{eq:swirlingMet} to the Levi-Civita metric is discussed in~\cite{Astorino:2022aam}. In the same work, a numeric treatment suggests that geodesic motion is vortex-like.\footnote{Recently, a very detailed analysis of the geodesics in the swirling background and in the exterior of the swirling black hole, has been carried out in~\cite{Capobianco:2023kse}.} The swirling spacetime is free of curvature singularities, a Misner string, and nonchronal regions. Finally, it is interesting to remark that the metric function $\omega$ grows infinitely large as $|z|\to\infty$. Being linear in $z$, it is constant on fixed-$z$ planes and zero on the equatorial plane $z=0$, where it changes sign. Do also note that this solution can be obtained from a Minkowski seed via a magnetic Ehlers transformation with parameter $j$.

\subsection{The geometry\label{sec:EMS}}

Let us then construct a new spacetime which features both, an external electromagnetic field and swirling rotation.\footnote{During the final stages of this work, we have noticed the thesis \cite{Illy:2023iau}. In there, an accelerated Reissner--Nordstr\"om black hole was constructed in a magnetic swirling background.}
We will create this from a Minkowski seed via a composition of magnetic Ehlers and Harrison transformations, in particular 
\begin{equation}
    \operatorname{E}[j]\circ\operatorname{H}[i\overline{X}],\label{eq:CompEHm}
\end{equation}
where the complex $X$ was defined directly below eq.~\eqref{eq:MelvinMet}. Since this is the first solution we present in this work, we will execute the steps listed in Sec.~\ref{sec:ErnstForma} one by one. Our seed metric is Minkowski in cylindrical coordinates, 
\begin{equation}
    ds_0^2 = -dt^2+d\rho^2+dz^2+\rho^2d\phi^2. 
\end{equation}
This can be cast into the metric~\eqref{eq:mWLP} with nonvanishing seed functions
\begin{equation}
    f_0=\rho^2=\mathrm{e}^{2\gamma}.
\end{equation}
Eqs.~\eqref{eq:eqchiM} and~\eqref{eq:eqATt} then yield vanishing twist potentials up to the choice of integration constants. The seed potentials are then the simplest possible, $\ernst_0=-\rho^2$ and $\Phi_0=0$.

We act upon them with the transformation~\eqref{eq:CompEHm} to obtain the new potentials
\begin{equation}
    (\ernst,\Phi)=\frac{(\ernst_0,i\overline{X}\ernst_0)}{1+(ij-|X|^2)\ernst_0},\label{eq:EhlersHarrisonTF}
\end{equation}
from which we may read off
\begin{equation}
    \begin{split}
        f&=\frac{\rho^2}{V^2+j^2\rho^4},\\
        \chi&=jf\rho^2,
    \end{split}
    \quad 
    \begin{split}
        A_\phi&=-\frac{f}{2}(BV-jE\rho^2),\\
        \widetilde{A}_t&= (A_\phi)_{(E,B)\to (-B,E)},
    \end{split}
\end{equation}
where $V$ was defined directly below~\eqref{eq:MelvinMet}, and $(A_\phi)_{(E,B)\to (-B,E)}$ denotes the value of $A_\phi$ with $B$ exchanged with $E$ and $E$ exchanged with $-B$. We use $|w|=\sqrt{w\overline{w}}$ for the modulus of a complex variable $w$. Plugging the above into eq.~\eqref{eq:eqchiM}, we obtain a pair of differential equations, first-order in derivatives of $\omega$, which we can integrate for 
\begin{equation}
    \omega = -4jz.
\end{equation}
With $\omega$ available, everything shall be fed to eq.~\eqref{eq:eqATt} which now yields a pair of differential equations, first-order in derivatives of $A_t$, the solution of which reads 
\begin{equation}
    A_t=-zf[EV^2\rho^{-2}+j(2BV-jE\rho^2)].
\end{equation}

It follows that the metric describing the \textit{electromagnetic swirling universe} (EMS) is 
\begin{equation}
    ds^2_{\mathsf{EMS}}=\frac{\rho^2}{V^2+j^2\rho^4}(d\phi+4jz\,dt)^2+(V^2+j^2\rho^4)(-dt^2+d\rho^2+dz^2),\label{eq:EMSmetric}
\end{equation}
accompanied by a gauge field 
\begin{equation}\label{eq:GFEMS}
    A=\rho^2 \frac{2z[EV^2\rho^{-2}+j(2BV-jE\rho^2)]dt+(BV-jE\rho^2)d\phi}{2(V^2+j^2\rho^4)}.
\end{equation}
It is straightforward to see that when the Harrison parameter vanishes, i.e., $X=0$ or equivalently, $E=0=B$, the gauge field vanishes and, taking into account that $V=1$ in such a case, we recover the swirling metric~\eqref{eq:swirlingMet}. On the other hand, when the Ehlers parameter vanishes, it is also easy to verify that the resulting spacetime is the electromagnetic universe with metric~\eqref{eq:MelvinMet} and gauge field~\eqref{eq:MelvinGF}. The metric~\eqref{eq:EMSmetric} admits a nonabelian group of motions $G_4$, locally generated by the KVs in the swirling case, eq.~\eqref{eq:swirlingKVs}. Of course, equality at the level of the algebras does not in general imply a group isomorphism (consider covering groups for example). In addition, we have a different Killing--Yano 2-form,
\begin{equation}
    -4j^2z\rho\,dt\wedge d\rho + (|X|^2V+j^2\rho^2)(V^2+j^2\rho^4)dt\wedge dz+j\rho d\rho\wedge d\phi,
\end{equation}
which reduces to the Killing--Yano 2-form~\eqref{eq:KYswirling} when we switch off the Harrison parameter (after a harmless overall division by $j$). On the other hand, if we make $j$ vanish, we get a Killing tensor $\propto -\partial_t\otimes \partial_t + \partial_z\otimes\partial_z$, which is just a trivial Killing tensor in the case of the electromagnetic universe. 

Let us now have a closer look at the metric~\eqref{eq:EMSmetric}. First of all, observe that the $\rho$ coordinate is not the so-called reduced circumference. The latter reads
\begin{equation}
    \mathsf{R}:=\frac{\sqrt{g_{\phi\phi}}}{2\pi}=\frac{\rho}{\sqrt{V^2+j^2\rho^4}},
\end{equation}
which goes to zero both as $\rho\to 0$ and $\rho\to\infty$. In fact, its maximum is at a $\rho_{\mathsf{max}} = (j^2+|X|^4)^{-1/4}$ with 
\begin{equation}
    0<\mathsf{R}\leq \mathsf{R}_{\mathsf{max}}:=\left(2|X|^2+2\sqrt{j^2+X^4}\right)^{-1/2}.
\end{equation}
Hence, this would make a very restricted coordinate, and this is why we will stick to the use of the initial $\rho$ coordinate. There are only two metric functions that change sign, $g_{tt}$ and $g_{t\phi}$. For the former, the surface where the change of sign happens, that is the surface on which $g_{tt}=0$, is given by the equation $\mathcal{S}=0$, where
\begin{equation}
    \mathcal{S}(\rho,z) := (V^2+j^2\rho^4)^2-(4jz\rho)^2,
\end{equation}
This actually defines two surfaces $\mathcal{S}_{\pm}=0$ (the + for positive $z$ and the $-$ for negative), with 
\begin{equation}
    \mathcal{S}_{\pm}:=V^2+j^2\rho^4\mp 4jz\rho,\label{eq:ergosurEMS}
\end{equation}
on which $\partial_t$ is null, and whose unit normals
\begin{equation}
    n_{\pm}^\mu = \frac{g^{\mu\nu}\partial_\nu \mathcal{S}_{\pm}}{\sqrt{g^{\lambda\sigma}(\partial_\lambda \mathcal{S}_{\pm})\partial_\sigma \mathcal{S}_{\pm}}},
\end{equation}
are spacelike, i.e., $n^\mu_{\pm} n^\nu_{\pm} g_{\mu\nu}\overset{\mathcal{S}_{\pm}}{=}1$. This means that the surfaces are timelike. 

Observe that $\mathcal{S}(\rho,-z)=\mathcal{S}(\rho,z)$, or, equivalently, that $\mathcal{S}_{\pm}(\rho,-z)=\mathcal{S}_{\mp}(\rho,z)$, and that $\mathcal{S}_{\pm}(\rho,0)\neq 0$. This implies that the equator, which is also the timelike surface where $g_{t\phi}=0$, acts as a plane of reflection, with $\mathcal{S}_-=0$ in the $z<0$ half-space being the mirror image of $\mathcal{S}_+=0$ in the positive $z$ half-space. Since 
\begin{equation}
    g_{tt}=-\frac{\mathcal{S}_+\mathcal{S}_-}{V^2+j^2\rho^4},
\end{equation}
it is clear that for $z>0$, $\mathcal{S}_->0$, and that if 
\begin{equation}
    z>\frac{V^2+j^2\rho^4}{4j\rho},
\end{equation}
then $g_{tt}>0$, viz. $\mathcal{S}_+<0$ gives a region in which $\partial_t$ is spacelike. Similarly, for $z<0$, $\mathcal{S}_+>0$, and if 
\begin{equation}
    z<-\frac{V^2+j^2\rho^4}{4j\rho},
\end{equation}
this provides another region, this time $\mathcal{S}_-<0$, in which $\partial_t$ is again spacelike. These two regions pretty much fulfill the criteria to be called \textit{ergoregions}, with $\mathcal{S}_{\pm}=0$ giving the \textit{ergosurfaces}. However, there might be a caveat with this interpretation, which requires further investigation and a deeper understanding. 

In the familiar Kerr geometry, $\partial_t$ can be selected as the unique timelike and normalized KV at infinity. Asymptotic flatness of a metric guarantees that the ergoregion (if it exists) is confined. Here, the metric exhibits a peculiar asymptotic behavior (it is basically asymptotically swirling as we will soon see). In fact, it is not hard to see that regions, where $\partial_t$ is spacelike, extend to infinity. Indeed, take $z$ to grow faster than $\rho^3$, and notice that the second term in~\eqref{eq:ergosurEMS} becomes dominant, yielding 
\begin{equation}
    g_{tt}\sim \frac{(4j\rho z)^2}{V^2+j^2\rho^4}>0.
\end{equation}
On the other hand, if $z$ grows slower than $\rho^3$, it is the first term in $\mathcal{S}_{\pm}$ that prevails, yielding 
\begin{equation}
    g_{tt}\sim -(j^2+|X|^4)\rho^4<0.
\end{equation}
Therefore, there are regions at infinity where $\partial_t$ is not timelike. Actually, there is simply no such KV (or a linear combination of them) in our case. To see this, consider the most general linear combination $\xi=\sum_{i=1}^4C_iT_i$, where the $T_i$'s are given in eq.~\eqref{eq:swirlingKVs}, with the $C_i$'s being constant coefficients. Being a linear combination of KVs with constant coefficients, this is obviously another KV. Fix a $t$, do $(\rho,z)\to(\rho\epsilon^{-1},z\epsilon^{-4})$, and Taylor expand about $\epsilon=0$. This ensures that we are probing a case where $z$ grows faster than $\rho^3$, as fast as $\rho^4$ in particular. Doing so, one confronts the following situation: it is impossible to choose the constants $\{C_i\}$ in a way such that the leading-order term in the expansion is negative! In other words, there is a region at asymptotic infinity where no KV can be timelike. Therefore, the concept of $t$ as a time of distant observers or a ``time at infinity'' seems to be somehow problematic, to say the least. 

Well, even in Kerr spacetime, it is true that the interpretation of $t$ as a ``time'', universal in the entirety of ``space'' (as a time of distant observers), is meaningful only outside the ergosphere~\cite{Frolov:1998wf}. Here, it just happens that the ergoregion unfortunately extends to infinity. Of course, it is always possible to find a KV which is timelike at $\rho$ infinity provided that $z$ grows slower than $\rho^3$. Truly, $\partial_t$ is such a KV, but so are other combinations $\xi$, e.g. $\xi=C_1T_1+C_2T_2+C_4T_4$ with $C_1^2>C_4^2$. Clearly, a normalization condition at infinity cannot be used here to single out a unique combination, for there is no $\xi$ satisfying this at all; indeed, $\xi\cdot\xi\sim -(C_1^2-C_4^2)(j^2+|X|^4)\rho^4$. One may however demand that $\xi\cdot\xi\sim -1$ as $\rho\to 0$ (after all this is a physical region). This forces $C_1^2=C_4^2+1$, but still leaves $C_2$ completely arbitrary. Moreover, the condition that $\xi\cdot\xi$ is time independent further fixes $C_4=0$, so we are left with $\xi=\partial_t + C_2\partial_\phi$. There is no other condition, based on limits, which we can use to somehow fix $C_2$. Note that one confronts the same situation also in the swirling spacetime. 

It is certainly tempting to consider the discrete symmetries of the metric as a means to fix $C_2$. Recall that in a Kerr geometry, reflection of time $t\to-t$ is not a symmetry unless it is accompanied by a change in the direction of rotation, namely $\phi\to-\phi$, and vice versa. This is true also for the EMS metric~\eqref{eq:EMSmetric}, except now we have additional ways to restore time-reversal symmetry. As a matter of fact, time reversal here, if accompanied by a transformation $z\to-z$ (which maps one semi-axis to the other), is another (simultaneous) discrete symmetry transformation; it leaves the metric invariant. Do also notice that a transformation $\phi\to-\phi$, again accompanied by $z\to-z$, is yet another discrete symmetry. These additional discrete symmetries are a key characteristic of the EMS and swirling metrics. They do not exist for example in the Kerr case. It then appears natural to ponder whether demanding that the norm of the KV candidate is invariant under all the aforementioned discrete symmetries of the metric, namely $(t,\phi)\to-(t,\phi)$, $(t,z)\to -(t,z)$, and $(z,\phi)\to -(z,\phi)$, is a good condition (on top of the previous ones), one that does the job. This indeed yields $C_2=0$, resulting in $\xi=\partial_t$. However, this does not prove uniqueness, although it naively appears to do so. 

To see this, consider the harmless coordinate transformation $(t',z',\phi')=(t,z-\alpha,\phi+4j\alpha t)$ where $\alpha$ is some real constant. The transformation is (metric-)form preserving, and therefore, the metric has the same discrete symmetries. Since the inner product $\xi\cdot \xi$ is a coordinate scalar, we may directly write it in terms of the prime coordinates. Now, $\partial_t\cdot\partial_t$ was invariant under $(t,z)\to-(t,z)$, but it will \textit{not} be invariant under $(t',z')\to-(t',z')$. However, the norm of another KV, namely $\partial_{t'}=\partial_t-4j\alpha\partial_\phi$, will. In fact, $\partial_{t'}$ also satisfies all the preceding criteria by default, and we see that discrete symmetries cannot help us single out a candidate KV after all. Our naively ``unique'' $\partial_t$ is as good as any other member of the family $\partial_t-4j\alpha\partial_\phi$. Therefore, in the absence of a robust selection mechanism, we argue that one should indeed use the whole family $\xi=\partial_t+C\partial_\phi$ as the timelike KV, where $C$ is an arbitrary real constant. It then follows that ergosurfaces would be understood as (the timelike) surfaces where $\xi$ is null, ergoregions as regions where $\xi$ is spacelike, and the frame-dragging angular velocity would be given by 
\begin{equation}
    \Omega := - \frac{\xi\cdot \partial_\phi}{g_{\phi\phi}}.\label{eq:FDAVoneparam}
\end{equation}

The latter deserves a few more comments. In particular, consider for a moment the general WLP metric in the form~\eqref{eq:mWLP}, for which we have the convenient equation
\begin{equation}
    \Omega=-C+\omega.
\end{equation}
It becomes evident, that using $\Omega$ to measure the value of the angular speed at each point is not really a meaningful practice, for $C$ is completely arbitrary. Instead, the meaningful quantity to look at is $\Delta\Omega(\rho,z):=\Omega(\rho,z)-\Omega_0=\omega(\rho,z)-\omega_0$. Remarkably, it then seems that for such spacetimes, in which the timelike KV is the one-parameter family $\xi$, rotation can only be understood in a relative manner. For example, in the EMS case we currently study, $\Omega=-C-4jz$. It is clear that, taking $\Omega_0$ to be the value of $\Omega$ on an arbitrary $z$ slice, this slice will be the surface where the difference $\Delta\Omega$ changes sign.\footnote{This will also act as the plane of reflection for the ergoregions.} This ultimately implies the presence of counter-rotating regions, though the exact localization of these regions is obviously observer-dependent. After this elucidating aside, let us now, \textit{once and for all}, choose coordinates adapted to a $C=0$ observer (our timelike KV being $\xi=\partial_t$), and let us proceed with discussing further features of the solution.

Going to a rectangular coordinate system, one can easily show that the metric has no coordinate singularities. There is no event horizon, and the absence of a conical singularity can be verified by the fact that 
\begin{equation}\label{eq:consing}
    \lim\limits_{\epsilon\to0}\frac{\int_0^{2\pi}\sqrt{g_{\phi\phi}}|_{\rho=\epsilon}\,d\phi}{\int_0^{\epsilon}\sqrt{g_{\rho\rho}}\,d\rho}=2\pi.
\end{equation}
The absence of a Misner string is also evident since the norm of the azimuthal KV vanishes as $\sim\rho^2$ near the symmetry axis. The electromagnetic swirling universe is also free of Closed Timelike Curves (CTCs), for $\partial_\phi$ is everywhere spacelike. Probing the spacetime for curvature singularities, we shall have a look at $R^{\mathbf{ab}}{}_{\mathbf{cd}}$, i.e., (the components of) the Riemann tensor in the orthonormal basis $\{e_{\mathbf{a}}\}$, defined in App.~\ref{app:ON&CNtet}. If this tensor is regular near a coordinate singularity, then the singularity is just due to a poor choice of chart. Indeed, all curvature invariants up to arbitrary polynomial order can be constructed using this particular tensor. Thus, if the tensor itself is regular, the regularity of the invariants follows. On the other hand, if $R^{\mathbf{ab}}{}_{\mathbf{cd}}$ is singular near a locus of interest, this does not automatically imply the existence of a curvature singularity, for the poles appearing in the tensor components could, in theory, not appear when taking traces to form curvature invariants. We find that $R^{\mathbf{ab}}{}_{\mathbf{cd}}$ depends solely on $\rho$, and that it is regular everywhere since the denominator of all components is just $(V^2+j^2\rho^4)^3$.\footnote{Recall that this goes to 1 when $\rho\to 0$.} We also verify that it falls off quite fast as $\rho\to\infty$, which reassures us that tidal forces are diminishing as one moves far away from the axis. To give an example, we mention that the Kretschmann scalar goes as $64(5|X|^4-3j^2)+\mathcal{O}(\rho)$ in the neighborhood of the symmetry axis, whereas it falls off as $\sim\rho^{-12}$ when $\rho\to\infty$.  

The new metric is asymptotic to 
\begin{equation}
    ds^2_{\mathsf{EMS}}\underset{\rho\to\infty}{\sim}\frac{1}{(|X|^4+j^2)\rho^2}(d\phi+4jz\,dt)^2+(|X|^4+j^2)\rho^4(-dt^2+d\rho^2+dz^2),
\end{equation}
provided that $z$ grows slower than $\rho^3$. A coordinate rescaling 
\begin{equation}
    \{t,\rho,z\} = \frac{|j|}{\sqrt{j^2+|X|^4}}\{t',\rho',z'\},\quad \phi = \frac{j^2+|X|^4}{j^2}\phi',
\end{equation}
brings the above to a form asymptotic to a swirling metric with parameter 
\begin{equation}
    j' = \frac{(j^2 + |X|^4)^2}{j^3}.
\end{equation}
However, the gauge field strength 2-form does not vanish as $\rho'\to\infty$. In fact, 
\begin{equation}
    F\underset{\rho'\to\infty}{\sim}\frac{j^2[(j^2-|X|^4)E-2jB|X|^2]}{(j^2+|X|^4)^2}dz'\wedge dt'
\end{equation}
Therefore, the complete solution is not asymptotic to the swirling spacetime because the latter is a vacuum solution. Note that $\ast (F\wedge\ast F)\to 0$ as $\rho'\to\infty$. 

Concerning the Petrov type of the EMS spacetime, it is straightforward to conclude that it is D, for we find that $9\Psi_2^2=\Psi_0\Psi_4$ (see the reasoning and other details in Sec.~\ref{sec:Petrov}). On the contrary, it is not trivial at all to prove that~\eqref{eq:EMSmetric} actually belongs to Kundt's class. Solutions in the Kundt family admit a shearfree, nonexpanding, and nontwisting null geodesic congruence, with the general metric being~\cite{Kundt1961,Kundt1962ExactSO,Stephani:2003tm,griffiths_podolsky_2009}
\begin{equation}
    ds^2 = 2 P^{-2}d\zeta\, d\overline{\zeta} -2du(dv+W\,d\zeta +\overline{W}\,d\overline{\zeta}+H\,du),\label{eq:KundtMetric}
\end{equation}
where $P,H$ are real functions, and $W$ is complex. Now, consider the specific functions 
\begin{equation}
P^2=\frac{\widetilde{z}^2+\gamma^2}{\widetilde{z}+k},\quad
        W=-\frac{\sqrt{2}v}{(\widetilde{z}+i\gamma)P^2},\quad
        H=\frac{\widetilde{z}^2+\gamma^2}{2}-\frac{2\gamma^2 v^2}{(\widetilde{z}^2+\gamma^2)^2P^2},\label{eq:KundtFuncs}
\end{equation}
where 
\begin{equation}\label{eq:gammaToj}
    \gamma = \frac{j}{2^{2/3}(j^2 +|X|^4)^{2/3}},\quad k = -\frac{\gamma X \overline{X}}{j},
\end{equation}
and where $\widetilde{z}$ is supposed to be given in terms of $\Re\zeta$ via $\sqrt{2}P^2d\widetilde{z}=d\zeta+d\overline{\zeta}$. Let us then perform the coordinate redefinitions
\begin{equation}\label{eq:KundtToNonExp}
\begin{split}
    \sqrt{2}\zeta &= \frac{\widetilde{z}(\widetilde{z}-2k)}{2}+(\gamma^2+k^2)\ln\frac{\widetilde{z}^2+\gamma^2}{P^2}+i(\psi+\gamma q^2),\\
    v&=q(\widetilde{z}^2+\gamma^2),\quad u=\tau-q,
    \end{split}
\end{equation}
in order to express~\eqref{eq:KundtMetric} in the coordinate system $\{\tau,\widetilde{z},q,\psi\}$. We get 
\begin{equation}
    ds^2 = (\widetilde{z}^2+\gamma^2)(-d\tau^2+dq^2)+P^{-2}(d\psi +2\gamma q\,d\tau)^2+P^2\,d\widetilde{z}^2,\label{eq:NonExpMet}
\end{equation}
which we readily recognize as a member of the general family of nonexpanding type D solutions (see (16.27) in~\cite{griffiths_podolsky_2009}). At this stage, yet another coordinate transformation with 
\begin{equation}\label{eq:NonExpToEMS}
    \begin{split}
        \tau&=t(\gamma^2+k^2)^{-1/2},\\
        \widetilde{z}&=-k+\rho^2(\gamma^2+k^2)^{-1}/4,
    \end{split}
    \quad 
    \begin{split}
        q&=z(\gamma^2+k^2)^{-1/2},\\
        \psi&=2\phi(\gamma^2+k^2),
    \end{split}
\end{equation}
finally brings us to the metric~\eqref{eq:EMSmetric}, and that is all. Therefore, we conclude that the EMS universe is also Kundt, as a combination of two Kundt spacetimes, the swirling one ($k=0$ via $X=0$) and the EM universe ($\gamma=0$ via $j=0$).

At the same time, there is a gauge field that we completely neglected so far. To find its form in coordinates $\{u,v,\zeta,\overline{\zeta}\}$, we start the other way around. Let 
\begin{equation}
    j=\frac{\gamma}{4(\gamma^2+k^2)^2},\quad X=\frac{\widetilde{E}+i\widetilde{B}}{2(\gamma^2+k^2)},\label{eq:FromCylToNonExpRedef}
\end{equation}
where $k=-\widetilde{E}^2-\widetilde{B}^2$, and perform the coordinate transformations 
\begin{equation}\label{eq:FromCylToNonExpCTs}
    \begin{split}
        t&=\tau\sqrt{\gamma^2+k^2},\\
        \rho&=2\sqrt{\gamma^2+k^2}\sqrt{\widetilde{z}+k},
    \end{split}
    \quad 
    \begin{split}
        z&=q\sqrt{\gamma^2+k^2},\\
        \phi&=\psi(\gamma^2+k^2)^{-1}/2,
    \end{split}
\end{equation}
to arrive at 
\begin{eqnarray}
    A&=&q\frac{2\gamma\widetilde{B}(\widetilde{z}+k)(k\widetilde{z}-\gamma^2) -\widetilde{E}[k^2\widetilde{z}^2 - (k^2+4k\widetilde{z}+\widetilde{z}^2)\gamma^2+\gamma^4]}{(\gamma^2+k^2)(\widetilde{z}^2+\gamma^2)}d\tau\nonumber\\
    &&+(\widetilde{z}+k)\frac{\gamma\widetilde{E}(\widetilde{z}+k)+\widetilde{B}(k\widetilde{z}-\gamma^2)}{(\gamma^2+k^2)(\widetilde{z}^2+\gamma^2)}d\psi,
\end{eqnarray}
which is the form of the gauge field~\eqref{eq:GFEMS} in the coordinate system $\{\tau,\widetilde{z},q,\psi\}$ with the spacetime metric being~\eqref{eq:NonExpMet}. At this stage, we shall consider 
\begin{equation}\label{eq:NonExpToKundt}
    \begin{split}
        \tau&=u+v/(\widetilde{z}^2+\gamma^2),\\
        q&=v/(\widetilde{z}^2+\gamma^2),
    \end{split}
    \quad
    \begin{split}
        \psi &= \sqrt{2}\Im\zeta - \gamma v^2/(\widetilde{z}^2+\gamma^2)^2,\\
        d\widetilde{z} &= (\widetilde{z}+k)(d\zeta + d\overline{\zeta})(\widetilde{z}^2+\gamma^2)^{-1}/\sqrt{2},
    \end{split}
\end{equation}
which bring us to 
\begin{eqnarray}
    A&=&v\frac{2\gamma \widetilde{B}(\widetilde{z}+k)(k\widetilde{z}-\gamma^2)-\widetilde{E}[k^2\widetilde{z}^2 - (k^2+4k\widetilde{z}+\widetilde{z}^2)\gamma^2+\gamma^4]}{(\gamma^2+k^2)(\widetilde{z}^2+\gamma^2)}du\nonumber\\
    &&-\frac{v\widetilde{E}}{(\widetilde{z}^2+\gamma^2)^2}dv+A_\zeta\,d\zeta + \overline{A}_{\zeta}\,d\overline{\zeta},
\end{eqnarray}
where 
\begin{eqnarray}
    A_\zeta&=&-i\widetilde{B}\frac{(\widetilde{z}+k)(k\widetilde{z}-\gamma^2)}{\sqrt{2}(\gamma^2+k^2)(\widetilde{z}^2+\gamma^2)}+\widetilde{E}(\widetilde{z}+k)\frac{2\widetilde{z}v^2(\gamma^2+k^2)-i\gamma(\widetilde{z}+k)(\widetilde{z}^2+\gamma^2)^3}{\sqrt{2}(\gamma^2+k^2)(\widetilde{z}^2+\gamma^2)^4},\nonumber\\
\end{eqnarray}
and where $\widetilde{z}$ is implicitly given in terms of $\Re\zeta$ via the second equation in the right column of~\eqref{eq:NonExpToKundt}. From here, one can check that using the backwards coordinate transformations~\eqref{eq:KundtToNonExp} and~\eqref{eq:NonExpToEMS}, together with~\eqref{eq:gammaToj} and 
\begin{equation}
    \widetilde{E}=E(\gamma^2+k^2),\quad \widetilde{B}=B(\gamma^2+k^2),
\end{equation}
one indeed reaches eq.~\eqref{eq:GFEMS}.

Finally, let us have a look at the electric and magnetic fields in the EMS universe. Following the procedure presented in App.~\ref{app:EMfields}, we find that 
\begin{equation}
    \mathbf{E}=-\frac{2jBV\rho^2+E(V^2-j^2\rho^4)}{(V^2+j^2\rho^4)^2}\hat{z},\quad \mathbf{B}=(\mathbf{E})_{(E,B)\to (-B,E)}.
\end{equation}
Both depend only on $\rho$ with field lines parallel to the axis of symmetry, in the vicinity of which they acquire a constant profile, $-E\hat{z}$ and $B\hat{z}$, respectively. Moreover, they fall off as $\sim \rho^{-4}$ when $\rho$ grows large, with the extrema of their magnitude given by the positive real roots of a hexic polynomial in $\rho$. 

\subsection{Double Wick rotation and the planar Reissner--Nordstr\"om--NUT spacetime}

It is a well-established result that the Bonnor-Melvin solution can be mapped to a planar Reissner--Nordstr\"om (RN) spacetime~\cite{Gibbons:2001sx}. It is also known that the swirling solution can be mapped to a planar Taub--NUT spacetime~\cite{Astorino:2022aam}. These mappings are in general achieved by employing a double Wick rotation, coordinate transformations, and parameter redefinitions. Therefore, one may reasonably expect to be able, in a similar fashion, to map the electromagnetic swirling universe, which utterly is a combination of the above, to a planar RN--NUT spacetime. Indeed, we prove that this is the case.

First, let us show how to ``planarize'' the standard RN--NUT metric via a limiting process. Here, the word ``standard'' refers to the metric
\begin{equation}
    d s^2_{\mathsf{RNN}} =-f(d \tau -2lx\,d\varphi)^2+f^{-1}{d r^2}+(r^2+l^2)[d x^2{(1-x^2)^{-1}}+(1-x^2)d \varphi^2],
\end{equation}
and the gauge field
\begin{equation}
    A_{\mathsf{RNN}} = \frac{gl-er}{r^2+l^2}d\tau + x\frac{2elr+g(r^2-l^2)}{r^2+l^2}d\varphi,\label{eq:RNNgf}
\end{equation}
where 
\begin{equation}
    f(r)=\frac{r^2-2m r-l^2+q^2}{r^2+l^2},\quad q^2=e^2+g^2,
\end{equation}
with $e$ and $g$ being the electric and magnetic charge parameters, respectively, $l$ denoting the NUT parameter, and $m$ standing for the mass parameter. 

We can ``planarize'' this solution, i.e., ``flatten'' the $S^2$ into $\mathbb{R}^2$, by performing the rescalings 
\begin{equation}
    {y}=\lambda \widetilde{y},\quad {r}=\widetilde{r}/\lambda,\quad {m}=\widetilde{m}/\lambda^3,\quad e= \widetilde{e}/\lambda^2,\quad g=\widetilde{g}/\lambda^2,\quad l=\widetilde{l}/\lambda,
\end{equation}
and sending $\lambda\to 0$ with $m,e,g,l\to\infty$ while $\widetilde{m},\ \widetilde{e},\ \widetilde{g}$, and $\widetilde{l}$ are kept fixed. Here, $y=\{\tau,x,\varphi\}$. This procedure results in the metric
    \begin{eqnarray}
         ds^2_{\mathsf{pRNN}} = -{f}(d {\tau}-2{l}x\,d \varphi)^2+f^{-1}{d r^2}+(r^2+l^2)(d x^2+d\varphi^2),\label{eq:pRNNmet}
    \end{eqnarray}
with 
\begin{equation}
    {f}=\frac{-2mr+q^2}{r^2+l^2},
\end{equation}
and in a gauge field equal to~\eqref{eq:RNNgf}. We have dropped the use of tilde accents for convenience. Notice that the gauge field is invariant under this process. Of course, the result is not necessarily guaranteed to be a solution; this is something that has to be checked. Nevertheless, one can verify that the full solution, comprised of the metric~\eqref{eq:pRNNmet} and the gauge field~\eqref{eq:RNNgf}, does indeed satisfy the Einstein--Maxwell field equations. Note that if $m>0$ and $r>0$, there is a Killing horizon at $r=q^2/(2m)$ (same as in the case of the planar RN solution), separating the inner region $0<r<q^2/(2m)$ from the outer one, where $t$ is timelike and $r$ is spacelike in the former and the other way around in the latter. On the other hand, if $m<0$ and $r>0$, and in contrast to the situation in the planar RN spacetime, the planar RN-NUT does not suffer from a naked singularity exactly due to the presence of the NUT parameter; the curvature scalars are everywhere regular. Of course, the planar RN--NUT solution is also plagued with a Misner string, for the symmetry axis cannot be well-behaved both at $x=1$ and $x=-1$.

Now, let us perform a double Wick rotation $(t,\phi)=i(\widetilde{\phi},\widetilde{t}\,)$ of~\eqref{eq:EMSmetric} and~\eqref{eq:GFEMS}, also doing $E=i\widetilde{E}$ and $B=i\widetilde{B}$ such that $\widetilde{X}:=\widetilde{E}+i\widetilde{B}=i X$. After the coordinate transformations 
\begin{equation}\label{eq:CTsTopRNN}
    \begin{split}
        \widetilde{t}&={4|m|^3}(4m^2l^2+q^4)^{-1}\tau,\\
        \rho &=\sqrt{4m^2l^2+q^4}\sqrt{-2mr+q^2}/(2m^2),
    \end{split}
    \quad 
    \begin{split}
        z&=x\sqrt{4m^2l^2+q^4}/(2|m|),\\
        \widetilde{\phi}&=\varphi\sqrt{4m^2l^2+q^4}/(2|m|),
    \end{split}
\end{equation}
and the parameter redefinitions
\begin{equation}
    j=-8l|m|^5(4m^2l^2+q^4)^{-2},\quad |\widetilde{X}|^2 = 4m^4q^2(4m^2l^2+q^4)^{-2},\label{eq:ParamDef1}
\end{equation}
we find out that the resulting metric is exactly the planar RNN metric~\eqref{eq:pRNNmet}. However, to bring the resulting gauge field into the form~\eqref{eq:RNNgf}, an additional gauge transformation is necessary, the purpose of which is to shift the temporal component by the specific constant $2m(2 glm-eq^2)(4m^2l^2+q^4)^{-1}$. Only then, the further parameter redefinitions
\begin{equation}
    \widetilde{E}=4m^2\frac{4elmq^2-g(4 m^2l^2-q^4)}{(4m^2l^2+q^4)^2},\quad \widetilde{B}=(|m|/m)(\widetilde{E})_{(e,g)\to (g,-e)},
\end{equation}
which satisfies the right equation in~\eqref{eq:ParamDef1}, leading to the desired result, namely the mapping of the gauge field in the EMS universe to~\eqref{eq:RNNgf}. Consequently, we conclude that the full solution can be consistently mapped to a planar RN--NUT spacetime via the above sequence of operations. The various limits are then clear. Killing $l$ is tantamount to switching off $j$, and vice versa; this provides the (bijective) mapping of the electromagnetic universe to the planar RN spacetime~\cite{Gibbons:2001sx}. Killing $e,g$ is tantamount to switching off $X$, and the other way around; this gives the mapping of the swirling solution to the planar Taub--NUT spacetime~\cite{Astorino:2022aam}. 

\subsection{Adding a cosmological constant}

The previous result motivates one to use the extension of the RN--NUT spacetime for a nonvanishing cosmological constant $\Lambda$, to derive a generalization of the EMS solution which includes a $\Lambda\neq 0$. In general, when a cosmological constant is included, the system of field equations is no longer integrable. Equations that were homogeneous in the absence of $\Lambda$, become inhomogeneous in its presence. In particular, the WLP metric itself is not suitable for stationary axisymmetric fields in the presence of a cosmological constant,\footnote{See~\cite{Charmousis:2006fx} for a generalized metric which, given a certain harmonic condition, can be reduced to the WLP metric.} \textit{ergo} the machinery used so far is not applicable. This is why the task of extending stationary and axisymmetric solutions of the Ernst equations to account for the presence of $\Lambda$, is a highly nontrivial one. 

Having said that, let us now attack the problem of generalizing the EMS solution. It can be straightforwardly checked that the form of the planar RN--NUT spacetime in the presence of $\Lambda$, is given by~\eqref{eq:pRNNmet} and~\eqref{eq:RNNgf}, with 
\begin{equation}
    f=\frac{-2mr+q^2+\Lambda l^4 - \Lambda r^2(r^2+6l^2)/3}{r^2+l^2}.
\end{equation}
Considering the inverse forms of the coordinate transformations~\eqref{eq:CTsTopRNN}, together with the parameter redefinitions 
\begin{equation}
    l=-\frac{j|m|^{1/3}}{2^{1/3}(j^2+|\widetilde{X}|^4)^{2/3}},\quad q^2=\frac{(2m^2)^{2/3}|\widetilde{X}|^2}{(j^2+|\widetilde{X}|^4)^{2/3}},
\end{equation}
and doing a double Wick rotation $(\widetilde{t},\widetilde{\phi})=i(\phi,t)$ of the resulting spacetime, also setting $\widetilde{X}=iX$, we obtain the generalized metric 
\begin{equation}
    ds^2_{\mathsf{EMS\Lambda}} = \frac{\mathcal{R}}{V^2+j^2\rho^4}(d\phi+4jz\,dt)^2 + (V^2+j^2\rho^4)\left(-dt^2 + \frac{\rho^2d\rho^2}{ \mathcal{R}}+dz^2\right),\label{eq:EMSLamMet}
\end{equation}
where 
\begin{eqnarray}
    \mathcal{R}(\rho)&=&\Lambda\frac{12j^4-(3j^2+|{X}|^4)^2}{12(j^2+|{X}|^4)^3}+\left(1-\frac{\Lambda(3j^2+|{X}|^4)|{X}|^2}{3(j^2+|X|^4)^2}\right)\rho^2\nonumber\\
    &&-\frac{\Lambda\rho^4}{2}\left(1 +\frac{2|X|^2}{3}\rho^2+\frac{j^2+|X|^4}{6}\rho^4\right).
\end{eqnarray}
Up to gauge transformations, the resulting gauge field is~\eqref{eq:GFEMS} as expected. Of course, there is no guarantee that the metric~\eqref{eq:EMSLamMet} and the gauge field~\eqref{eq:GFEMS} solve the Einstein--Maxwell-$\Lambda$ field equations, but we verify that this is the case indeed. Therefore, we have successfully constructed the \textit{cosmological extension} of the EMS spacetime, which also is of Petrov type D, as well as a member of the Kundt class.\footnote{Applying the parameter redefinitions~\eqref{eq:FromCylToNonExpRedef} and the coordinate transformations~\eqref{eq:FromCylToNonExpCTs}, one should be readily convinced that the transformed metric belongs to the general family of nonexpanding type D solutions. From there, reaching the Kundt form is more or less straightforward (see~\cite{griffiths_podolsky_2009}).}

Observe that for $\partial_\phi$ to remain spacelike at infinity, we must consider $\Lambda<0$. We also see that there is a spinning string at $\rho=0$, for 
\begin{equation}
    \lim\limits_{\rho\to 0}g_{\phi\phi}=\Lambda\frac{3j^4-|X|^4(|X|^4+6j^2)}{12(j^2+|X|^4)^3},
\end{equation}
Let us then do $(t,\phi)=(\widetilde{t}+a\widetilde{\phi},\widetilde{\phi})$, with $a$ being a constant, together with a reidentification of the new coordinates such that $\widetilde{\phi}$ is $2\pi$ periodic, to see whether we can obtain a new spacetime free of it. It is not difficult to check that $g_{\widetilde{\phi}\widetilde{\phi}}$ has a $z$ dependence, and that the axis can be made regular only at a single $z$ (by fixing $a$), meaning that the spinning string will persist. Therefore, we deduce that it is not artificial (it cannot be removed everywhere along the $z$ axis); it rather corresponds to an actual Misner string! Fortunately, we can remove it by tuning the Ehlers and Harrison parameters as
\begin{equation}
    j^2 = {\frac{3+2\sqrt{3}}{3}}|X|^4,\label{eq:TuneMisner}
\end{equation}
before any regluing. Doing so, we then observe that the induced metric with ${t}=\mathrm{cte}=z$, after the optional rescaling $\rho= \widetilde{\rho}\sqrt{4|X|^2-\Lambda}/(2|X|)$, assumes the form 
\begin{equation}
    d{s}^2_{\mathsf{EMS\Lambda}}|_{t,z} \underset{\widetilde{\rho}\to 0}{\sim} d\widetilde{\rho}^2 + \frac{(4|X|^2-\Lambda)^2}{16|X|^4}\widetilde{\rho}^2d{\phi}^2 ,\label{eq:NearOrigAngu}
\end{equation}
close to the symmetry axis.\footnote{If we do not rescale $\rho$, then eq.~\eqref{eq:NearOrigAngu} will be the same up to multiplication by a constant factor. This will not change the value of the deficit, since it does not alter the ratio between the proper length of a circumference and the radius.} The above expression suggests the presence of an infinite strut with negative mass per unit length
\begin{equation}
    \mu=\delta/4=\Lambda/(4|X|^2),
\end{equation}
where $2\pi\delta$ is the excess angle of the line source. Thankfully, this can be made to vanish via the rescaling 
\begin{equation}
    \phi=\frac{4|X|^2}{4|X|^2-\Lambda}\widetilde{\phi},
\end{equation}
if we reidentify $\widetilde{\phi}$ as our new azimuthal coordinate with period $2\pi$. Consequently, it is always possible to obtain a regular spacetime with a negative cosmological constant, free of a Misner string, conical singularities, and CTCs, with the caveat of having the particular relation~\eqref{eq:TuneMisner} between $j$ and $X$. 

\begin{figure}
    \centering
    \includegraphics{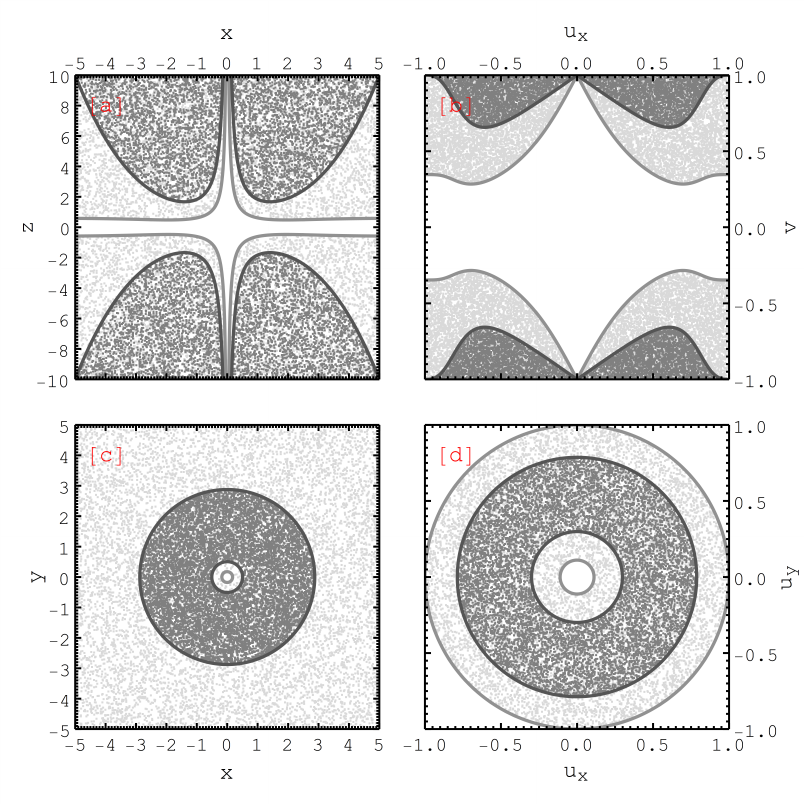}
    \caption{Cross-sections of the ergoregions in the EMS (stippled dark gray areas) and EMS-$\Lambda$ (stippled light gray areas) spacetimes. Plots are for $|X|=0.35$, and are given by eq.~\eqref{eq:TuneMisner}. The cosmological constant in the EMS solution has been set to $-3$. In panel [a] we display a ${\mathrm{y}}=0$ cross-section with rectangular coordinates $\mathrm{(x,y)}=\rho(\sin\phi,\cos\phi)$. In panel [c] we display a $|z|=3$ cross-section. Similarly, in panel [b], a $u_{\mathrm{y}}=0$ slice is displayed using coordinates $(u_{\mathrm{x}},u_{\mathrm{y}})=(2/\pi)(\operatorname{atan}\rho)(\sin\phi,\cos\phi)$ and $v=(2/\pi)\operatorname{atan} z$, where $0\leq u_{\mathrm{x}}^2+u_{\mathrm{y}}^2<1$. Finally, in panel [d] we show a $|v|=(2/\pi)\operatorname{atan}3$ plane section. Coordinates $\{u_{\mathrm{x}},u_{\mathrm{y}},v\}$ are particularly convenient, since radial infinity is approached via $u_{\mathrm{x}}^2+u_{\mathrm{y}}^2\to1$ and $|z|$ infinity via $|v|\to 1$, with the symmetry axis located at $u_{\mathrm{x}}=0=u_{\mathrm{y}}$.}
    \label{fig:EMSΛErgo}
\end{figure}

After imposing the tuning~\eqref{eq:TuneMisner} by replacing $j$, and with $\widetilde{\phi}$ being our new azimuthal coordinate, the $\mathcal{R}$ function in the metric~\eqref{eq:EMSLamMet} acquires the form 
\begin{equation}
    \mathcal{R}=\rho^2\frac{-9\Lambda+2|X|^2[18-\Lambda\rho^2(9+6|X|^2\rho^2+(3+\sqrt{3})|X|^4\rho^4]}{36|X|^2}.
\end{equation}
This is a positive function, with the reduced circumference being (dropping the tilde accent)
\begin{equation}
    \mathsf{R}:=\frac{\int_{0}^{2\pi}\sqrt{g_{{\phi}{\phi}}}\,d{\phi}}{2\pi}=\frac{4|X|^2\sqrt{\mathcal{R}}}{(4|X|^2-\Lambda)\sqrt{V^2+{\frac{3+2\sqrt{3}}{3}}|X|^4\rho^4}}.
\end{equation}
In contrast to $\mathsf{R}$ in the case $\Lambda=0$, here the reduced circumference, being a monotonically increasing function of $\rho$, has the same range as the $\rho$ coordinate. Moreover, the presence of a negative cosmological constant has a significant impact on the full extent of the ergoregions as we see in the instructive Fig.~\ref{fig:EMSΛErgo}. 
Indeed, if we let both $\rho$ and $|z|$ approach infinity, we find that 
\begin{equation}
    g_{tt}\underset{\rho,|z|\to\infty}{\sim} - \frac{4j^2\Lambda z^2\rho^4}{3}>0,
\end{equation}
where $j,X$ are understood as being related via eq.~\eqref{eq:TuneMisner}. In fact, as $\rho\to\infty$, it suffices that $|z|\gtrapprox 1.05/\sqrt{-\Lambda}$ for $\partial_t$ to be spacelike. Close to the axis, that is as $\rho\to 0$, we have 
\begin{equation}
    g_{tt}\overset{|z|\in\mathcal{O}(\rho^{-2})}{\underset{\rho,|z|^{-1}\to 0}{\sim}}\propto (4|X|^2-\Lambda)\rho^2 z^2>0,
\end{equation}
where we let $z$ approach infinity at least as fast as $\rho^{-2}$. As in the EMS case, the ergoregions do not touch the $z$ axis. 

Finally, for completeness, let us mention that the metric~\eqref{eq:EMSLamMet} with $\Lambda>0$, after taking care of the Misner string by imposing eq.~\eqref{eq:TuneMisner}, features a cosmological horizon located at the largest positive real root of a polynomial hexic in $\rho$. This can be utterly written as a polynomial cubic in $\rho^2$ which, using Descartes' rule of signs, appears to have a single positive real root if and only if $X^2>\Lambda/4$. This is then a double root of the hexic equation. Alas, the whole situation (with positive $\Lambda$) gets more complicated, if we notice that as we approach this root, call it $\rho_+$, the reduced circumference vanishes, meaning that $\rho=\rho_+$ behaves as a sort of ``axis'' besides $\rho=0$. This observation is rather expected, simply because the root of $g^{\rho\rho}$ is necessarily a root of $g_{\phi\phi}$, as can be seen from the line element~\eqref{eq:EMSLamMet}.

Let us also briefly go through the various limits. When $X=0$, the gauge field vanishes, and the cosmological EMS metric~\eqref{eq:EMSLamMet} reduces to the cosmological extension of the swirling solution presented in~\cite{Astorino:2022aam}. The latter is free of curvature singularities and free of any horizons if $\Lambda<0$. Notwithstanding these good features, a fact completely neglected in~\cite{Astorino:2022aam}, is that the cosmological swirling solution actually features a cosmic spinning string, evident from 
\begin{equation}
    \lim\limits_{\rho\to 0} g_{\phi\phi}=\frac{\Lambda}{4j^2},
\end{equation}
which proves to be irremovable through coordinate transformations and regluing. As such, it shall again be understood as a Misner string. Since we no longer have the freedom to tune parameters in order to remove it (as we did previously), one comes to the unfortunate conclusion that the swirling-$\Lambda$ solution of~\cite{Astorino:2022aam} is in general plagued with a Misner string and the CTCs accompanying it.

When $j=0$, the metric~\eqref{eq:EMSLamMet} acquires the static form
\begin{equation}
    ds^2 = \frac{\rho^2 -\Lambda |X|^{-4}V^4/12}{V^2}d\phi^2+V^2\left(-dt^2+\frac{\rho^2d\rho^2}{\rho^2 -\Lambda |X|^{-4}V^4/12} +dz^2\right),
\end{equation}
and the gauge field becomes~\eqref{eq:MelvinGF}. We remark that this is not the metric presented in the perhaps pertinent cases~\cite{Zofka:2019yfa,Astorino:2012zm}. The above spacetime seems to have a spinning string, for 
\begin{equation}
    \lim\limits_{\rho\to 0}g_{\phi\phi}=-\Lambda|X|^{-4}/12.
\end{equation}
If $\Lambda>0$, it is impossible to remove this string by any means. If on the other hand $\Lambda<0$, we can do 
\begin{equation}
    (t,\phi)=\left(\widetilde{t}+\frac{\sqrt{-\Lambda}}{2\sqrt{3}X\overline{X}}\widetilde{\phi},\widetilde{\phi}\right)
\end{equation}
and reglue the spacetime to get rid of it. The new spacetime is then also free of conical singularities. An expansion of the induced metric with $\widetilde{t}=\mathrm{cte}=z$ near the symmetry axis attests to that. Moreover, $\partial_{\widetilde{\phi}}$ is clearly spacelike everywhere if we adhere to the use of a negative cosmological constant, while we also remark that there are no curvature singularities, nor are any ergoregions present. 

\subsection{Embedding a Schwarzschild black hole}

Having analyzed the background, it is time to discuss the EMS black hole which is just the Schwarzschild black hole embedded in the previously discussed spacetime. For this reason, we will refer to it also as \textit{Schwarzschild--EMS}. This embedding shall be understood as a composition of magnetic Ehlers and Harrison transformations acting on the potentials associated with a Schwarzschild spacetime. We will work with the spherical-like coordinates $\{t,r,x=\cos\theta,\phi\}$,\footnote{Here, the coordinate $x$ ought not to be confused with the usual Cartesian coordinate $\mathrm{x}$ used in other parts of this work.} since these prove to be the most convenient for integration. 

It is quite straightforward to carry out the first steps which result in identifying the Schwarzschild metric 
\begin{equation}
    ds^2_{\mathcal{S}}=-(1-2M/r)dt^2+(1-2M/r)^{-1}dr^2+r^2(1-x^2)^{-1}dx^2+r^2(1-x^2)d\phi^2,
\end{equation}
with the magnetic WLP metric~\eqref{eq:mWLP}, the nonvanishing metric functions being 
\begin{equation}\label{eq:EMSSseed}
    f_0=r^2(1-x^2),\quad \mathrm{e}^{2\gamma}=\frac{r^4(1-x^2)}{(r-M)^2-M^2x^2},
\end{equation}
with Weyl's coordinates $\rho,z$ given in terms of $r,x$ via 
\begin{equation}
    \rho=\sqrt{r(r-2M)(1-x^2)},\quad z=(r-M)x.\label{eq:RhoZ}
\end{equation}
Since this is a static vacuum solution, it is evident that $\Phi_0=0$ and $\ernst_0=-f_0$. Observe that the mass does not appear in the potentials, but rather in the function $\gamma$ and the coordinate transformations~\eqref{eq:RhoZ}.

Having retrieved the seed data, we now act with the transformation~\eqref{eq:EhlersHarrisonTF} on the seed potentials to obtain---after integrating the twist equations---a new spacetime, comprised of the metric
\begin{equation}\label{eq:ScEMSmet}
    ds^2_{\mathsf{\mathcal{S}EMS}}=f(d\phi-\omega\, dt)^2+(1-x^2)f^{-1}r^2ds^2_{\mathcal{S}}|_{d\phi=0},
\end{equation}
with functions
\begin{equation}
    \begin{split}
        f&=\frac{r^2(1-x^2)}{V^2+j^2r^4(1-x^2)^2},\\
        \omega&=-4j(r-2M)x,
    \end{split}
\end{equation}
together with the gauge field
\begin{eqnarray}
    A&=&-(r-2M)x\frac{E[V^2-j^2r^4(1-x^2)^2]+2jBr^2(1-x^2)V}{V^2+j^2 r^4(1-x^2)^2}dt\nonumber\\
    &&+\frac{r^2(1-x^2)}{2}\frac{jE r^2(1-x^2)-B V}{V^2+j^2 r^4(1-x^2)^2}d\phi,
\end{eqnarray}
where the function $V$ is now defined as 
\begin{equation}
    V(r,x):=1+|X|^2r^2(1-x^2).
\end{equation}

This spacetime describes the exterior of a black hole with a Killing horizon, located at $r_+=2M$, dressing a singularity at the radial origin. Indeed, $g^{rr}(r_+,x)=0$, and the Kretschmann scalar blows up as $\sim M^2 r^{-6}$ only when $r\to 0$. The surface area of the horizon is that of a 2-sphere of radius $r_+$, namely $4\pi r_+^2$. However, its circumference at the equator $x=0$ is not $2\pi r_+$ in the presence of the transformation parameters, i.e., 
\begin{equation}
    \int_0^{2\pi}\sqrt{g_{\phi\phi}(r=r_+,x=0)}\,d\phi = \frac{2\pi r_+}{\sqrt{(1+|X|^2 r_+^2)^2+j^2 r_+^4}}.
\end{equation}
This suggests that the horizon surface, its embedding in particular, can be visualized as a 2-sphere of area $4\pi r_+^2$, which is deformed when we switch on $j$ and/or $X$. In particular, the deformation caused by these parameters can be visualized as squeezing the 2-sphere on the equatorial plane, progressively obtaining an egg-shaped surface which is deformed into a peanut-like one as we squeeze stronger. Moreover, the solution is stationary; the black hole gets dragged due to the swirling property of the background. It also got ``electromagnetized'', in the sense that it is no longer a vacuum solution, but rather an electrovac one, a fact ascribed to the presence of the external electric and magnetic fields in the background geometry. Note that~\eqref{eq:ScEMSmet} enjoys the good features of the EMS background, i.e., it is free of topological singularities and nonchronal regions,\footnote{The reader may convince herself/himself by quickly performing the pertinent checks we previously did in the case of the background.} while it also shares the same asymptotic behavior with the latter. 

\begin{figure}
    \centering
    \includegraphics{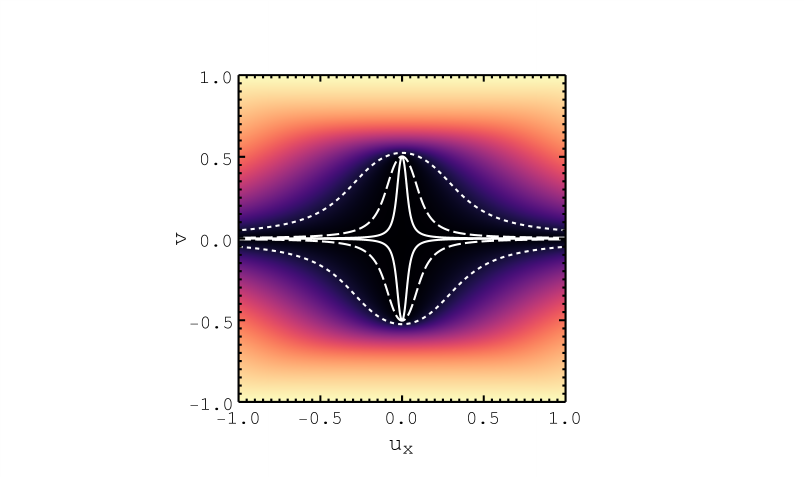}
    \caption{Heatmap of $\widetilde{\omega}:=(2/\pi)\operatorname{atan}|\omega(u_{\mathrm{x}},v)|$ on the $u_{\mathrm{x}}v$-plane for $j=0.5$ and $M=1$. Lightest hue corresponds to $\widetilde{\omega}=1$, whereas darkest hue (pure black) to $\widetilde{\omega}=0$. The darker the hue, the smaller $\widetilde{\omega}$ is. The horizon should be understood as the $[-1/2,1/2]$ part of the $v$ axis. Note that $\widetilde{\omega}$ vanishes only on the equatorial plane and in the horizon limit. We also display the following isolines: $\widetilde{\omega}=10^{-3}$ (solid), $\widetilde{\omega}=10^{-2}$ (dashed), and $\widetilde{\omega}=0.1$ (dotted).}
    \label{fig:EMSSergoOmega}
\end{figure}

Concerning $\omega$, we display various isolines over a heatmap in Fig.~\ref{fig:EMSSergoOmega} (see caption for details). We do so in coordinates $\{u_{\mathrm{x}},u_{\mathrm{y}},v\}$, introduced previously. 
In fact, we first switch to coordinates~\eqref{eq:RhoZ} via the inverse transformations
\begin{equation}
   \begin{split}
       r&=M+\sqrt{\frac{\rho^2+z^2+M^2+\sqrt{(\rho^2+z^2+M^2)^2-4M^2 z^2}}{2}},\\
       x&=\frac{\sqrt{2}z}{\sqrt{\rho^2+z^2+M^2+\sqrt{(\rho^2+z^2+M^2)^2-4M^2 z^2}}}.
   \end{split}
\end{equation}
In this chart, the horizon surface is understood as a closed line segment of length $2M$ on the $z$ axis (with center at $z=0$). We can formally approach it for $|z|\leq M$ by taking $\rho\to 0$. Then, we once again employ the convenient coordinates 
\begin{equation}
    u_{\mathrm{x}}=(2/\pi)(\operatorname{atan} \rho)\sin\phi,\quad u_{\mathrm{y}}=(2/\pi)(\operatorname{atan} \rho)\cos\phi,\quad v=(2/\pi)\operatorname{atan} z\label{eq:uvcoords}
\end{equation}
in which $|z|$ infinity sits at $|v|=1$ and $\rho$ infinity at $u_{\mathrm{x}}^2+u_{\mathrm{y}}^2=1$. The horizon is again understood as a line segment, now of length $(4/\pi)\operatorname{atan} M$, on the $v$ axis (with center at $v=0$). 

In Fig.~\ref{fig:EMSSergoOmega} we (indirectly) see that $|\omega|$ grows infinitely large as $|z|\to\infty$. Since it describes the frame-dragging angular velocity, the fact that it is linear in $z$, implies that the two $z$ half-spaces (positive and negative) counterrotate. We also observe that it gets smaller and smaller as we approach the equator, where it vanishes. It also decreases as we approach the horizon where it also vanishes in the respective limit. Note that, for $j>0$, $\omega<0$ in the half-space defined by $z>0$ (or $v>0$), positive otherwise; the exact opposite holds true when $j<0$. Regarding the ergoregions, these prove to be more or less insensitive to the introduction of a mass. Actually, their behavior at asymptotic infinity is exactly the same as that in the case of the EMS solution we previously discussed, and there is nothing, really worth reporting, going on close to the horizon.

Attacking the Petrov classification next, and following the reasoning presented in Sec.~\eqref{sec:Petrov}, we can deduce that the general Petrov type is I, for we have that $9\Psi_2^2\neq \Psi_0\Psi_4$. In particular, 
\begin{eqnarray}
    9\Psi_2^2- \Psi_0\Psi_4&=&18M(r-2M)(1-x^2)(j-i|X|^2)\nonumber\\
    &&\times\frac{\{jr^2(1-x^2)+i[1-|X|^2r^2(1-x^2)]\}^2[j r^2(1-x^2)+i V]^3}{r^4[V^2+j^2r^4(1-x^2)^2]^5}.
\end{eqnarray}
This becomes zero when $M=0$, or $j=0=X$, or at the poles $x=1$ and $x=-1$, or at the horizon $r=2M$. In the limit of vanishing mass, we recover the EMS universe (discussed in the previous sections), albeit in the spherical-like coordinate system $\{t,r,x,\phi\}$. When $j=0=X$, we obtain the Schwarzschild seed which is type D everywhere. At the poles, we observe that $\Psi_4\neq 0$, meaning that the Petrov type is D on the axis (at least in the exterior). On the horizon surface, the spacetime is algebraically special. Since $r=2M$ turns out to be a pole of $\Psi_4$, it would be erroneous to make a definite claim that the Petrov type is D; we can only argue (as we just did) that the solution is algebraically special, because $9\Psi_2^2- \Psi_0\Psi_4$,\footnote{Note that $9\Psi_2^2- \Psi_0\Psi_4=\sqrt{(I^3-27J^2)/(\Psi_0\Psi_4)}$ with $I, J$ as defined in~\cite{Stephani:2003tm}.} which vanishes at $r=2M$, utterly corresponds to a combination of curvature invariants and thus, it is independent of the reference frame. Note that our choice~\eqref{eq:ONTmag} of the orthonormal tetrad, from which we constructed the CNT (see App.~\ref{app:ON&CNtet} for details), corresponds to a zero angular momentum observer in circular motion; it is not suitable for studying the horizon limit. To determine the actual Petrov type at $r=2M$, one must rather consider a falling observer~\cite{Pravda:2005uv}. Finally, when $X=0$, we obtain the swirling black hole, the Petrov type of which is also I, whereas for $j=0$ we recover the static type I Schwarzschild--Melvin spacetime~\cite{ErnstSchwMel}, which describes a Schwarzschild black hole embedded into an (electro)magnetic universe. The latter solution is of Petrov type II on the horizon surface (again located at $r=2M$), and it also features another locus of interest given by 
\begin{equation}
    |X|\sqrt{1-x^2}=r^{-1},\quad x\neq 1,-1,
\end{equation}
where all Weyl scalars vanish and the Petrov type is, therefore, O~\cite{Pravda:2005uv}.

\section{Mixing electric and magnetic transformations}
\label{mixing}

In this section, we wish to scrutinize another possible route. Instead of composing transformations of the same ``kind'', i.e., either electric or magnetic, we shall explore their mixture. Once again, we will consider only Ehlers and Harrison transformations. In general, there are eight possible mixed compositions, 
\begin{equation}
    U^\alpha{}_\beta\circ U^\gamma{}_\delta,
\end{equation}
with $\beta\neq \delta$, where $\alpha,\ldots,\delta=1,2$ and 
\begin{equation}
    (U^\alpha{}_\beta)=\begin{pmatrix} \operatorname{E}_e  & \operatorname{E}_m \\ \operatorname{H}_e & \operatorname{H}_m\end{pmatrix}.
\end{equation}
However, if the seed is Minkowski, the number of available compositions reduces to four, namely
\begin{equation}
    U^\alpha{}_{1}\circ U^\beta{}_2.
\end{equation}
This happens because 
\begin{equation}
    U^{\alpha}{}_{2}\circ U^{\beta}{}_1 \sim U^{\alpha}{}_2,
\end{equation}
where $\sim$ denotes a rough equivalence relation here, in the sense that electric transformations of the seed potentials associated with Minkowski space, result in a metric which is also Minkowski modulo coordinate transformations. 

Our first course of action is to study the novel spacetimes obtained via these mixed compositions in the case of a Minkowski seed. For starters, we wish to see if these geometries, presented here for the first time, turn out to be backgrounds, i.e., that they are first and foremost free of curvature singularities, topological defects, and other sorts of pathologies. Since we know that $U^{1}{}_\alpha\circ U^2{}_{\alpha}=U^{2}{}_\alpha\circ U^1{}_{\alpha}$, we could directly operate on the seed potentials with $\operatorname{E}_e[c]\circ\operatorname{H}_e[Q]\circ\operatorname{E}_m[j]\circ\operatorname{H}_m[i\overline{X}]$, where $j,c$ are real parameters, $X$ was previously introduced in the case of the electromagnetic universe, namely $X=(E+iB)/2$, and we also define a new complex parameter $Q:=(q_e+i q_m)/2$. This would be tantamount to casting the EMS metric in the form~\eqref{eq:eleWLP} and acting on the associated electric potentials with $\operatorname{E}_e[c]\circ\operatorname{H}_e[Q]$. We could then consider all possible limits, in which we remain with two layers of transformations, one magnetic and one electric. Nevertheless, computationally speaking, this would not be a wise strategy to pursue. For this reason, we will generate each case separately. We will probe the new geometries only for curvature singularities, spinning strings, conical singularities, and nonchronal regions (regions with CTCs). We will call them backgrounds if they are free of curvature and topological singularities, proper backgrounds if they are also free of CTCs/CNCs.

\subsection{Electromagnetic universe and electric Harrison transformations\label{sec:Mix}}
Here, we operate on the magnetic potentials $\ernst_0=-\rho^2$ and $\Phi_0=0$, associated with Minkowski spacetime, with $\operatorname{H}_e[Q]\circ \operatorname{H}_m[i\overline{X}]$.\footnote{See p. 3 of this manuscript for nomenclature.} Of course, we do not have to truly do the composition; we may directly act on the electric potentials, associated with the electromagnetic universe, with $\operatorname{H}_e[Q]$. Hence, our seed potentials are 
\begin{equation}\label{seedbackpot}
    \ernst_0 = V^2-4|X|^2z^2,\quad \Phi_0 = -2\overline{X}z,
\end{equation}
where we recall that $V=1+|X|^2\rho^2$. We thus follow the prescription presented at the end of Sec.~\ref{sec:ErnstForma}, skipping the integration details to get the target metric~\eqref{eq:eleWLP}, with 
\begin{equation}\label{eq:MetricFuncsMelCha}
\begin{split}
        f&=\frac{V^2}{\mathcal{V}^2+2\mathcal{V}(q_eE - q_m B)z+16|QX|^2z^2},\\
        \omega&=C_\omega - (q_e B + q_m E)\frac{1+|Q|^2(V^2+4 |X|^4 \rho^2 z^2)}{|X|^2V} ,\\
        \mathrm{e}^{2\gamma}&=V^4,
\end{split}
\end{equation}
where we further defined $\mathcal{V}(\rho,z):=1-|Q|^2\ernst_0$ for convenience. Soon, we will also display the gauge field; before that, let us study this new metric which has Petrov type I. 

The first thing that one observes, is that on the equatorial plane $z=0$, $f$ has a double (and single) pole at $\rho=\rho_*$, where $\rho_*$ denotes the positive real root of $\mathcal{V}=0$, or $|Q|^2 V^2=1$ equivalently. We find that 
\begin{equation}
    \rho_* = \frac{\sqrt{1-|Q|}}{|X|\sqrt{|Q|}},
\end{equation}
which is clearly real if-f $|Q|\leq 1$. Saturating the bound, this coordinate singularity is put exactly at $\rho=0=z$. For $Q<1$, the locus is a ring. For $q_e B\neq -q_m E$, this is the only pole of $f$. In any case, if this is a true curvature singularity (be it a point or a ring), and since $f\neq 0$ for all $\rho,z$, indicating the absence of a horizon, it must be that it is a naked one, with the general consensus being that such configurations are unphysical. Note that the metric does not have any other singular limits besides the one we just reported, which unfortunately happens to be a curvature singularity, for we find that the Kretschmann scalar at the equator blows up as $\sim (\rho-\rho_*)^{-8}$ in the limit $\rho\to\rho_*$. Parameter tuning cannot be a remedy to this; there are simply not enough parameters to tune in order to get rid of all poles up to octic order in the pertinent Taylor expansion. The situation is worse when $|Q|=1$, in which case $R^{\mu\nu}{}_{\rho\sigma}R^{\rho\sigma}{}_{\mu\nu}\sim \rho^{-16}$ when $\rho\to 0$. Consequently, it is mandatory to restrict $|Q|>1$ in order to expel $\rho_*$ from the physical range of $\rho$. Still, regularity (of curvature invariants) is guaranteed only if the components of the Riemann tensor in the orthonormal basis of App.~\ref{app:ON&CNtet} are everywhere regular. Fortunately, we find that their denominators are proportional to $f^{-1}$, with the proportionality factors being nonvanishing functions of $\rho$. Since $f^{-1}=0$ has no positive real solutions for $|Q|>1$, it follows that the spacetime is indeed everywhere regular. 

Having successfully tackled this important issue, it is time to address another subtlety. Observe that at fixed $z$,
\begin{equation}
    \lim\limits_{\rho\to 0} g_{\phi\phi} = -\left(C_\omega - \frac{(1+|Q|^2)(q_eB+q_mE)}{|X|^2}\right)^2\lim\limits_{\rho\to 0}f,
\end{equation}
where the limit of $f$ as $\rho\to 0$ is a nonvanishing expression that involves the parameters and $z$, appearing as a denominator of the leading term in the above. Once again, we are confronted with a spinning string, which we can remove by fixing the integration constant $C_\omega$ as 
\begin{equation}
    C_\omega = \frac{(1+|Q|^2)(q_eB+q_mE)}{|X|^2}.\label{eq:Comfix}
\end{equation}
Of course, keeping the integration constant free in this case, only to fix it now, was after all a proactive action. Had we set it to zero, we would again have the string removed via a coordinate transformation, followed by a regluing of spacetime. Nevertheless, the final expression becomes
\begin{equation}
    \omega = \frac{(q_eB+q_mE)[1-|Q|^2(V+4|X|^2 z^2)]\rho^2}{V},
\end{equation}
and we now also have well-defined limits $X\to 0$ (Minkowski), or $Q\to 0$ (electromagnetic universe).\footnote{Observe that, if $C_\omega$ is not fixed as in eq.~\eqref{eq:Comfix}, $X\to 0$ is a singular limit of $\omega$ in~\eqref{eq:MetricFuncsMelCha}.} Additionally, we remark that after taking care of the string, the induced metric with $t=\mathrm{cte}=z$ becomes $\sim C(d\rho^2+\rho^2 d\phi^2)$ in the vicinity of the symmetry axis, where $C$ denotes a proportionality factor depending on the parameters and the chosen value of $z$. This tells us that there are no conical singularities to be bothered with. Henceforth, we consider only the family with $|Q|>1$ and $C_\omega$ as in eq.~\eqref{eq:Comfix}.

\begin{figure}
    \centering
    \includegraphics{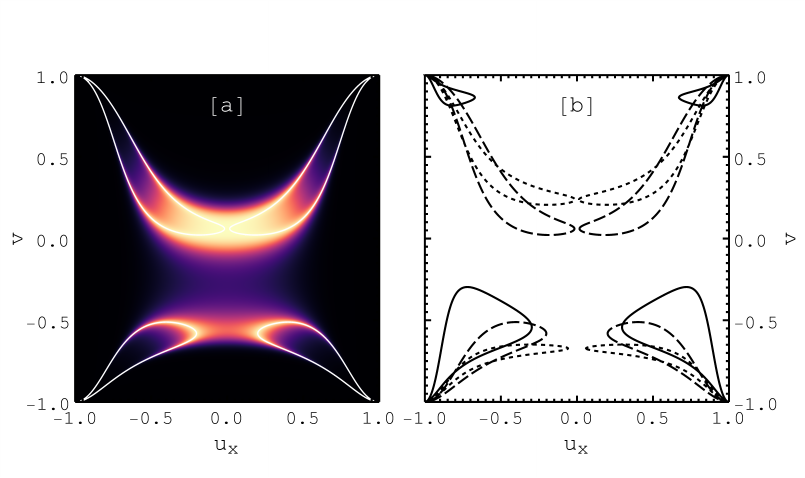}
    \caption{In panel [a] we plot a heatmap of $(2/\pi)\operatorname{atan}|\widetilde{\omega}|$ on the $u_{\mathrm{x}v}$-plane for $q_e=1.05$, $q_m=2.06$, $E=1$, and $B=-1$. We do so in coordinates~\eqref{eq:uvcoords}. The lighter the hue, the greater $|\widetilde{\omega}|$ is, with white denoting the singular surface. Regions enclosed by the white curves are regions of negative $\widetilde{\omega}$. In panel [b] we display a $u_\mathrm{y}=0$ cross-section of the $g_{\phi\phi}=0$ surface for $q_e=2.07$, $q_m=2.33$, $E=-0.31$, $B=0.06$ (solid), $q_e=1.05$, $q_m=2.06$, $E=1$, $B=-1$ (dashed), $q_e=0.01$, $q_m=-3.14$, $E=-0.07$, $B=0.92$ (dotted). Regions bounded by the curves are nonchronal regions (filled with CTCs). Keep in mind that the full picture is obtained via a complete revolution of these profiles about the $v$ axis, whereby infinity is depicted as a cylinder of unit height and radius.}
    \label{fig:CTCsEMC}
\end{figure}

For $|Q|>1$, the solution is everywhere stationary (not just static) provided that $q_e B\neq -q_m E$. If $q_eB+q_mE>0$, we have that $\omega<0$. If the former is negative, the latter is positive. It is also clear that as $\rho\to 0$, it holds that $\omega\to 0$. However, the metric function $\omega$ is no longer the object of interest here, because the frame-dragging velocity (in coordinates adapted to $\partial_t$) is actually given by 
\begin{equation}\label{eq:FDAV}
    \widetilde{\omega} = -\frac{\omega f^2}{\rho^2 - (\omega f)^2}.
\end{equation}
This function is too lengthy to write it down explicitly. However, it is obvious that its zeroes are the zeroes of $\omega$ (given that $f$ has no poles), its poles the zeroes of
\begin{equation}
     g_{\phi\phi} = \frac{\rho^2-(f\omega)^2}{f},
\end{equation}
i.e., the surface where the reduced circumference shrinks to zero. Since $\omega$ has a fixed sign depending on the choice of $q_eB+q_mE$, and since $f>0$ everywhere, it follows that the sign of $\widetilde{\omega}$ may change if and only if $g_{\phi\phi}$ changes sign, namely if there exist CTCs in this spacetime. The boundary of these nonchronal regions, that is the surface on which the norm of $\partial_\phi$ vanishes, will then be a surface of infinite $|\widetilde{\omega}|$. These observations are manifest in Fig.~\ref{fig:CTCsEMC}.
Indeed, observe that the white curves in panel [a] are exactly the dashed ones in panel [b]. Note that appearances can be deceiving here, for it looks like the nonchronal regions extend to infinity. This is not true; by doing $\rho\to\rho/\lambda$, $z\to z/\lambda^\alpha$ with $\alpha$ real positive, and expanding about $\lambda=0$, one may check that the leading term is always positive, whichever the value of $\alpha$ is. Interestingly, the fact that the chronology horizons~\cite{SHchrono} (the surfaces where the nonchronal regions meet the chronal ones) coincide with the surfaces of singular frame-dragging angular velocity, perhaps admits a physical interpretation. Rotation becomes very rapid very close to these singular surfaces, thereby dragging inertial frames so strongly that the light cones are completely tilted in the direction of the circumference! Such a situation is not unfamiliar generally speaking. A similar interpretation, roughly speaking, appears, for example, in the case of the Van Stockum solution~\cite{vanStockum:1937zz,BonnorDust1}.

Now, observe that if we tune our transformation parameters such that $q_mE =- q_e B$, $\omega$ vanishes and thus we obtain a static metric, which also is of Petrov type I. The new metric is now free of CTCs, because $g_{\phi\phi}=\rho^2/f$ with $f>0$ everywhere; there is no rotation taking place any longer to tilt the light cones. However, our claim that the spacetime is regular is not valid under the particular tuning. Indeed, the assumptions for that were $q_mE \neq - q_e B$ and $|Q|>1$. Unfortunately, when $q_mE =- q_e B$, there are two spatial surfaces, one for $z<0$ and the other for positive $z$, on which the curvature invariants become singular. The explicit surface equations are found by requesting the vanishing of the denominator of $f$. Therefore, although we got rid of rotation and the CTCs, we ended up with a far worse situation, namely a naked singularity with a weird disconnected geometry. It goes without saying that there is no reason to further discuss this scenario, and one should stick to the previous assumptions which at least guarantee regularity. 

\begin{figure}
    \centering
    \includegraphics{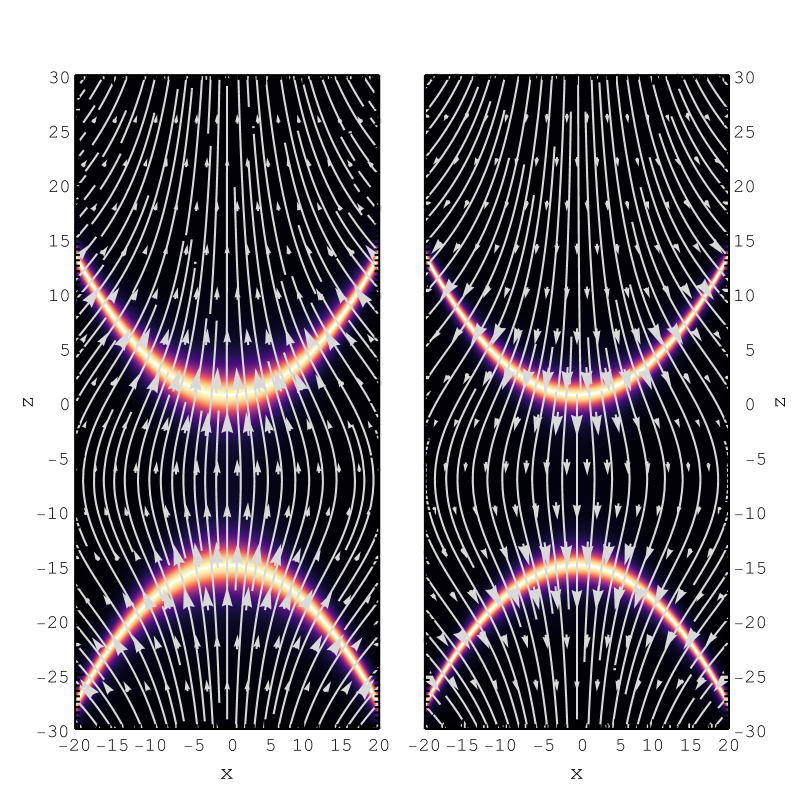}
    \caption{Field lines plotted over the heatmap of (a function of) the norm for the electric field (left panel) and the magnetic field (right panel). The size of the arrows is proportional to the magnitude of the vector at each sampled point. The heatmaps show the values of $\operatorname{atan}|\mathbf{E}|$ (left) and  $\operatorname{atan}|\mathbf{B}|$ (right). The lighter the color, the larger the number. The norms are everywhere finite, and the $\operatorname{atan}$ function is used for display purposes. Plots are for $q_e=-2$, $q_m=131/151$, $E=-2/17$, and $B=-131/2567$. These values satisfy eq.~\eqref{eq:ProperTune}.}
    \label{fig:EBfieldsEMC}
\end{figure}

Let us finally have a look at the gauge field. Its components read
\begin{equation}\label{eq:gaugeFuncsMelchar}
\begin{split}
         A_t&=f\frac{[\mathcal{V}+(q_eE-q_mB)z](-Ez+{q_e}\ernst_0/2) -(Bz+{q_m}\ernst_0/2)(q_eB+q_mE)z}{V^2},\\
        A_\phi&=\rho^2\frac{4|X|^2q_mz+(V+4|X|^2z^2)[({3q_e^2}-{q_m^2})B/4+q_eq_m E]-B}{2V} -\omega A_t,
\end{split}
\end{equation}
modulo gauge transformations. A visual of the field lines (and more) can be found in Fig.~\ref{fig:EBfieldsEMC}.
We mention that the electric and magnetic fields vanish at asymptotic infinity in all directions. Contrasting this with the behavior of the fields in the electromagnetic universe, in which they are uniform close to the axis for all $z$, we can argue that the electric and magnetic fields in this spacetime are better-behaved, at least in terms of asymptotic behavior. 

\subsection{Electromagnetic universe and electric Ehlers transformations}
Next, let us discuss an alternative possibility that generates a new type I axisymmetric stationary electrovac field, starting again with the electromagnetic universe as our seed. This scenario involves acting with $\operatorname{E}[c]$, where $c$ is a real parameter, upon the seed potentials~\eqref{seedbackpot}. Skipping the integration details (we just follow the algorithmic process outlined at the end of Sec.~\ref{sec:ErnstForma}), we obtain the target metric~\eqref{eq:eleWLP}, with functions 
\begin{equation}\label{eq:MetricFuncsMelNUT}
\begin{split}
        f&=\frac{V^2}{1+c^2(V^2-4|X|^2z^2)^2},\\
        \omega&=8c\frac{|X|^2\rho^2 z}{V} ,\\
        \mathrm{e}^{2\gamma}&=V^4,
\end{split}
\end{equation}
together with a gauge field, whose components read
\begin{equation}\label{eq:gaugeFuncsMelNUT}
\begin{split}
         A_t&=\frac{-E+cB(V^2-4|X|^2z^2)}{1+c^2(V^2-4|X|^2z^2)^2}z,\\
        A_\phi&=-\frac{B+cE(V+4|X|^2z^2)}{2V}\rho^2-\omega A_t.
\end{split}
\end{equation}

It is easy to see that if $c\neq 0$, the Petrov type is I. Of course, for $c=0$ the solution reduces to the electromagnetic universe which has Petrov type D. The metric does not exhibit any coordinate singularities, and the axis is not plagued with a spinning string. In particular, the induced metric with $t=\mathrm{cte}=z$ behaves as 
\begin{equation}
   ds^2 \underset{\rho\to 0}{\sim} [1+c^2(1-4|X|^2z^2)^2](d\rho^2+\rho^2d\phi^2),
\end{equation}
near the symmetry axis, which also proves that there is no conical singularity there. The components $R^{\mathbf{ab}}{}_{\mathbf{cd}}$ of the Riemann tensor in the pertinent orthonormal basis of App.~\ref{app:ON&CNtet}, have a denominator of the general form $V^n f^{-k}$, where $n$ and $k$ are irrelevant positive integers. Since this denominator cannot be made to vanish, we conclude that curvature invariants up to arbitrary polynomial order will be everywhere regular. Moreover, $\lim_{\rho\to\infty} R^{\mathbf{ab}}{}_{\mathbf{cd}}=0$ (ditto for $z\to\infty$) further ensures that tidal forces vanish as we move far away from the $z$ axis and/or the equator. We may then call this geometry a background, although not a proper one.

Note that $f$ is strictly positive continuous, meaning that there are no ergoregions in this stationary spacetime. However, there are surfaces where the reduced circumference shrinks to zero. These are, once again, exactly the surfaces where the frame-dragging angular velocity~\eqref{eq:FDAV} blows up. Actually, there are two separate $g_{\phi\phi}=0$ surfaces, one in each $z$ half-space, which moreover behave as chronology horizons, in the sense that they separate CTC-free regions from CTC-full ones. Therefore, the interpretation we gave in the previous section applies also here.\footnote{One can check that the nonchronal regions do not extend to infinity and that they exist for arbitrary values of $X$ and $c$.} Their structure is more or less similar to the one in the previous case, although now the equator functions as a plane of reflection, simply because $g_{\phi\phi}$ is invariant under $z\to-z$. For this reason, we do not bother plotting them.  

Now, regarding the electric and magnetic fields, the electric Ehlers transformation has---besides modifying the $z$ component---generated a nontrivial component in the $\hat{\rho}$ direction which, as expected, is proportional to $c$. Interestingly, there are surfaces where the radial component of the electric field (with respect to the unit basis) vanishes. These are given by 
\begin{equation}
    z^2 = \frac{c(-E+cBV^2)\pm2|c||X|}{4Bc^2|X|^2},
\end{equation}
provided that the right hand side is positive. On these surfaces,
\begin{equation}
    \mathbf{E} = \frac{B^2(2|c||X|\pm cE)}{2V^2[4E|c||X|-c(B^2+2E^2)]}\hat{z}=:CV^{-2}\hat{z},
\end{equation}
which is exactly the form of the electric (or magnetic) field in the electromagnetic universe! Note that both, electric and magnetic, fields vanish as $\rho\to\infty$. This was also the case in the seed spacetime. Remarkably, however, here they also vanish far away in the $\hat{z}$ direction, which was not at all the case in the Bonnor--Melvin solution, where the fields did not depend on $z$. In particular, we have that 
\begin{equation}
    \mathbf{E}+i\mathbf{B}\underset{z\to\pm\infty}{\sim}\frac{1}{4c|X|^2V}\left( \frac{1}{z^2V}\hat{z}\mp\frac{\rho}{z^3}\hat{\rho}\right)(B+iE).
\end{equation}

\subsection{Swirling universe and electric Ehlers transformations}
Having explored the scenarios with an electromagnetic universe as our seed, we shall now discuss our options when considering a swirling seed. Let us initiate this discussion with the following composition. In theory, we start with Minkowski space and operate with $\operatorname{E}_e[c]\circ \operatorname{E}_m[j]$ on the associated seed potentials. In practice, we will just act with $\operatorname{E}[c]$ upon the seed potentials associated with the swirling spacetime cast into the electric WLP form. Consequently, it is necessary to identify the seed quantities anew; the swirling spacetime can be described by a metric~\eqref{eq:eleWLP} with functions
\begin{align}
f_0&=\frac{S^2-(4j\rho z)^2}{S},\\
\omega_0&=\frac{4j\rho^2 z}{S^2-(4j\rho z)^2},\\
e^{2\gamma}&=S^2-(4j\rho z)^2,
\end{align}
where we have defined $S(\rho):=1+j^2\rho^4$. Given the fact that the seed is stationary, the solution to~\eqref{eq:eqchiE} is a nontrivial seed potential
\begin{equation}
\chi_0=-2j\left(\rho^2+6z^2-\frac{8z^2}{S}\right).
\end{equation}
With the previous quantities at hand, the only nonvanishing seed Ernst potential is found to be
\begin{align}
\ernst_0&=f_0+i\chi_0.\label{eq:SwiSeedEle}
\end{align}

After acting with $\operatorname{E}[c]$, we arrive at the metric~\eqref{eq:eleWLP} with functions
\begin{equation}\label{eq:MetricFuncsSwirNUT}
\begin{split}
        f&=\frac{f_0}{(1-c\chi_0)^2+c^2f_0^2},\\
        \omega&=\omega_0\left\lbrace1+2jc[(S+2)\rho^2-4z^2]-c^2[4-8j^2z^2\boldsymbol{(}2z^2+(S-2)\rho^2\boldsymbol{)}-3S^2]\right\rbrace.
\end{split}
\end{equation}
The former is directly read off from the target potential $\ernst$ since it corresponds to $\Re \ernst$ in the absence of $\Phi$. The latter requires integrating eq.~\eqref{eq:eqchiE}. Concerning the Petrov type, here we can explicitly write down the relevant expression because it is fairly short,
\begin{equation}\label{eq:DetPetrov}
    9\Psi_2^2 - \Psi_0\Psi_4 = -\frac{144cj^3\rho^2}{\left\lbrace 1+cj[(S+2)\rho^2-4z^2]+i[c-j\rho^2-cj^2\rho^2(12z^2+\rho^2)]\right\rbrace^5}
\end{equation}
in particular. Therefore, the Petrov type---based on the analysis we did in Sec.~\ref{sec:Petrov}---of the solution is I. The swirling universe is obtained in the limit $c\to 0$. In this case, expression~\eqref{eq:DetPetrov} vanishes, but $\Psi_4\neq 0$. On the other hand, in the limit $j\to 0$, all five complex Weyl-NP scalars become zero, indicating a Petrov type O. Indeed, the solution reduces to flat spacetime. This totally agrees with the fact that electric transformations, when applied to the potentials of Minkowski spacetime, give rise to a target spacetime which is again Minkowski modulo coordinate rescalings. For the case at hand, these rescalings read
\begin{equation}
t\rightarrow\frac{t}{\sqrt{1+c^2}},\quad (\rho,z) \rightarrow\sqrt{1+c^2} (\rho,z).
\end{equation}

The new metric is free of a spinning string, for $\lim_{\rho\to 0} g_{\phi\phi} =0$ at fixed arbitrary $z$. Moreover, it is also free of conical singularities; the induced metric with $t=\mathrm{cte}=z$ behaves as $\sim C(\rho^2+\rho^2 d\phi^2)$ near the symmetry axis, where $C$ is a constant depending on the parameters and the fixed value of $z$. Note that there are no coordinate singularities evident. However, the components of the Riemann tensor in the orthonormal basis, i.e., $R^{\mathbf{a}\mathbf{b}}{}_{\mathbf{c}\mathbf{d}}$, come with a denominator of the general form $[S^2-(4j\rho z)^2]^m[(1-c\chi_0)^2+c^2f_0^2]^n S^n$, where the exact values of the integers $m\geq0<n$ are utterly unimportant at this stage. When $m=0$ (and for some components it is), the denominator can never vanish in the admissible coordinate range. On the other hand, when $m\neq 0$ (true for some components), there are potential poles on the surfaces $\mathcal{S}_{\pm}=0$ (the ergosurfaces as we will soon see), with $\mathcal{S}_{\pm}$ given in eq.~\eqref{eq:ergosurEMS} for $X=0$. Therefore, one cannot argue that the spacetime is everywhere regular, at least not in the fashion we previously did; one needs to compute curvature invariants explicitly. Such behavior is solely due to the swirling nature of the target spacetime, for $\mathcal{S}_{\pm}$ is independent of $c$. Nevertheless, calculating the Kretschmann scalar, one finds that the denominator of the latter has $m=0$ and $n=6$, meaning that it is everywhere regular. Cubic polynomials also have $m=0$ ($n$=9). Consequently, at least up to third-order curvature polynomials (which are coordinate scalars), the absence of singularities is verified. The appearance of poles in higher order polynomials is highly unlikely then, although regularity is not guaranteed in the robust sense of having an everywhere regular $R^{\mathbf{ab}}{}_{\mathbf{cd}}$. Therefore, we may consider this as a background geometry, although we will immediately see that it cannot be proper.  

\begin{figure}
    \centering
    \includegraphics{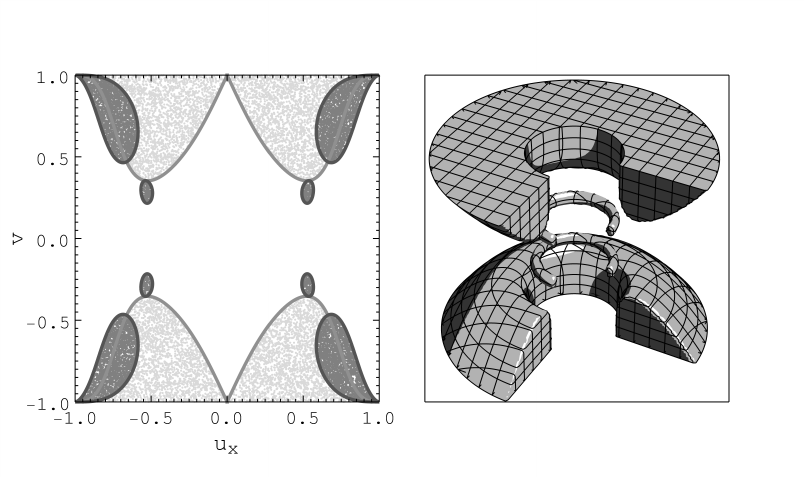}
    \caption{\textit{Left panel}. Showing $u_{\mathrm{y}}=0$ cross-section of the ergoregions (light gray) and the nonchronal regions (dark gray). \textit{Right panel}. 3D Illustration of the nonchronal regions (regions containing CTCs) in rectangular coordinates $\{\mathrm{x}=\rho\sin\phi,\mathrm{y}=\rho\cos\phi,z\}$. Plots are for $j=0.5$ and $c=-0.75$.}
    \label{fig:case3CTCs}
\end{figure}

Note that the denominator of $f$ is everywhere positive, meaning that the sign of $g_{tt}=-f$ only depends on the numerator. Its vanishing happens on loci satisfying the surface equation 
\begin{equation}
    \mathcal{S}_{+}\mathcal{S}_- = (1+j^2\rho^4)^2-(4jz\rho)^2=0,\label{eq:ErgoCase3}
\end{equation}
which gives the ergosurfaces. Consequently, the ergosurfaces in this spacetime are exactly the same as the ones in the swirling universe. Finally, let us once again probe for CTCs. By now, the narrative should be clear. For the metric~\eqref{eq:eleWLP}, the frame-dragging angular velocity is given by~\eqref{eq:FDAV}. Rotation becomes infinitely rapid on the surfaces where the denominator vanishes, provided that the zeroes of the denominator are not zeroes of the numerator, or if they are, that the denominator grows faster than $\omega f^2$ close to the surface. Then, these surfaces are necessarily zeroes of $g_{\phi\phi}$, and if $g_{\phi\phi}$ changes sign there, these act as chronology horizons, separating the chronal from the nonchronal parts of spacetime. In the previous examples of this section, the denominator of $g_{\phi\phi}$, namely the function $f$, was strictly positive. Thus, the sign of $g_{\phi\phi}$ solely depended on the sign of the numerator $\rho^2-(\omega f)^2$. Here, because there are ergoregions, both the numerator and the denominator are allowed to change sign. In fact, we expect that the ergoregions and the nonchronal regions containing CTCs, partially overlap, namely that chronology horizons and ergosurfaces cross each other. Indeed, this can be seen in Fig.~\ref{fig:case3CTCs}.  
Quite interestingly, it turns out that in the spacetime under study, there can actually be up to four disconnected regions filled with CTCs for certain parameter ranges, two toroidal regions with finite volume, and two other regions that extend to infinity, as can be seen from 
\begin{equation}
    g_{\phi\phi}\underset{\rho,|z|\to\infty}{\sim}-\frac{1024(cj\rho^2 z^3)^2}{144z^4+j^2\rho^8}<0,
\end{equation}
where we let $z$ grow exactly as fast as $\rho^2$. Do note that although it seems that the nonchronal ``tori'' closer to the equator comes into contact with the ergosurfaces, this is not the case. There are only two \textit{rings} where the two surfaces intersect. Looking at the left panel of Fig.~\ref{fig:case3CTCs}, these would be at $|v|\approx 0.47$ with radius $|u_{\mathrm{x}}|\approx 0.66$.

\subsection{Swirling universe and electric Harrison transformations}
The remaining spacetime to consider involves the action of the composition $\operatorname{H}_e[Q]\circ \operatorname{E}_m[j]$ on the seed potentials of Minkowski spacetime. In practice, we just act with $\operatorname{H}[Q]$ on the seed potentials~\eqref{eq:SwiSeedEle}. The integration details are ``left as an exercise''; the whole process is already described in the introduction of this work. We obtain the target metric~\eqref{eq:eleWLP} with functions
\begin{equation}\label{eq:MetricFuncsSwircharged}
\begin{split}
     f&=\frac{f_0}{(1-|Q|^2f_0)^2+|Q|^4\chi_0^2},\\
    \omega&=\omega_0\left[1-|Q|^4\left(4-8j^2z^2(2z^2+(S-2)\rho^2)-3S^2\right)\right].
\end{split}
\end{equation}
By doing this transformation, we have further excited a gauge field with nonvanishing components
\begin{equation}\label{eq:gaugeFuncsSwircharged}
\begin{split}
        A_t&=\frac{f}{2f_0}\left\lbrace q_e[f_0(1-|Q|^2f_0)-|Q|^2\chi_0^2] - q_m\chi_0 \right\rbrace,\\
    A_\phi&=\frac{\omega_0}{2}\left\lbrace q_mj[(S+2)\rho^2-4z^2]+[1-(\omega/\omega_0)(q_e/|Q|^2)]-\omega A_t \right\rbrace.
    \end{split}
\end{equation}

One can immediately check that $9\Psi_2^2\neq \Psi_0\Psi_4$ with $\Psi_4\neq 0$ for $j,Q\neq 0$. Therefore, the Petrov type of the solution is I. The swirling universe solution is recovered in the limit $Q\to 0$ (recall that this implies $q_e,q_m\to 0$). Indeed, also $9\Psi_2^2 = \Psi_0\Psi_4$ in this case, with $\Psi_4\neq 0$, which gives the Petrov type we expect, that is D. Minkowski spacetime (up to irrelevant coordinate rescalings) is obtained in the limit $j\to 0$. In this limit, all Weyl--NP scalars vanish and the type is O as expected. This spacetime is also free of a spinning string. The fact that $g_{\phi\phi}$ vanishes as $\sim \rho^2$ near the symmetry axis, proves the claim. It is also free of conical singularities, for the induced metric with $t=\mathrm{cte}=z$ behaves as $\sim C(d\rho^2+\rho^2d\phi^2)$ in the vicinity of the axis, with $C$ being a constant depending on the parameters and the fixed value of $z$. What about coordinate singularities?

Let us focus on $f$ and probe it for poles. The function is expected to blow up at solutions to the equation $(1-|Q|^2f_0)^2+|Q|^4\chi_0^2=0$. This is an equation quadratic in $z^2$, which does not have any real solution unless $z=0$. On the equatorial plane, the form 
\begin{equation}
    (1-|Q|^2)^2 + 2 j^2 |Q|^2(3|Q|^2-1)\hat{\rho} + j^4|Q|^4\hat{\rho}^2=0,\label{eq:PolesCase4}
\end{equation}
is assumed, where $\hat{\rho}:=\rho^{1/4}$. It is clear that the above admits only a single positive real solution if-f $|Q|=1$, the solution being $\rho=0$. This localizes the coordinate singularity to a single point $\rho=0=z$ (the origin). We then need to check whether if this is an honest curvature singularity, or just due to a poor choice of coordinates. Once again, we turn to $R^{\mathbf{ab}}{}_{\mathbf{cd}}$ in order to draw conclusions. In the case under study, the components have a denominator of the general form $[S^2-(4j\rho z)^2]^m[(1-|Q|^2f_0)^2+|Q|^4\chi_0^2]^n S^n$.\footnote{Again, the exact values of $m\geq 0<n$ are not needed to make the argument.} Clearly, regularity cannot be directly deduced from $R^{\mathbf{ab}}{}_{\mathbf{cd}}$, because this guy simply blows up on surfaces~\eqref{eq:ErgoCase3} and at the origin $\rho,z= 0$. This is not problematic per se, as long as the invariants are regular. Upon examining the behavior of the Kretschmann scalar for example, it becomes evident that the thing can be singular only at the origin, which it approaches as $\sim \rho^{-12}$, and only when $|Q|=1$. Consequently, if $|Q|=1$, the singularity sitting at $\rho=0=z$ constitutes a true curvature singularity. However, there is absolutely no null hypersurface to act as a sort of event horizon here, and one thus fails to comply with the censorship hypothesis; the singularity is naked. Moreover, the zeroes of the function $f$, i.e., the surfaces~\eqref{eq:ErgoCase3}, which represent the ergosurfaces in this spacetime, are completely independent of $Q$ (they coincide with the ergosurfaces in the swirling universe), meaning that they will also exist for $|Q|=1$. A naked singularity with ergoregions sounds indeed like things really took the wrong turn, and they most probably did. However, such a situation is not that unfamiliar, with the case of the hyperextreme Kerr black hole first coming to mind~\cite{griffiths_podolsky_2009}. As previously mentioned, naked singularities are generally regarded as unphysical. Fortunately, all we have to do to avoid this worrying issue, is to simply exclude the value $|Q|= 1$ from the pool of allowed parameter values. Doing so, we can then argue that spacetime is regular, of course in the less robust sense of the foregoing section, namely that curvature invariants at least up to some low order are regular everywhere. Then, since we have a geometry free of all sorts of ``physical'' singularities, we can again call this a background.

It turns out that this background cannot get the ``proper'' attribute, because, once again, we are confronted with CTCs. The nonchronal regions in this spacetime are more or less similar to the ones in the previous case (see Fig.~\ref{fig:case3CTCs}); there can be up to four disjoint regions, two in the $z>0$ half-space and their mirror images in the negative half-space, with the equator acting as the plane of reflection. As before, closer to the equator we have compact toroidal regions full of CTCs which exist if $|Q|>1$ regardless of the value of $j$. As $|z|$ grows larger, one enters the other nonchronal regions which again extend to infinity, as can be seen from 
\begin{equation}
    g_{\phi\phi}\underset{\rho,|z|\to\infty}{\sim} -\frac{1024(j|Q|^2\rho^2z^3)^2}{144 z^4 +j^2\rho^8}<0,
\end{equation}
where we take $z$ to grow exactly as fast as $\rho^2$. The two regions in each half-space merge exactly when $|Q|=1$, which was previously excluded to avoid the curvature singularity at the origin. 

\begin{figure}
    \centering
    \includegraphics{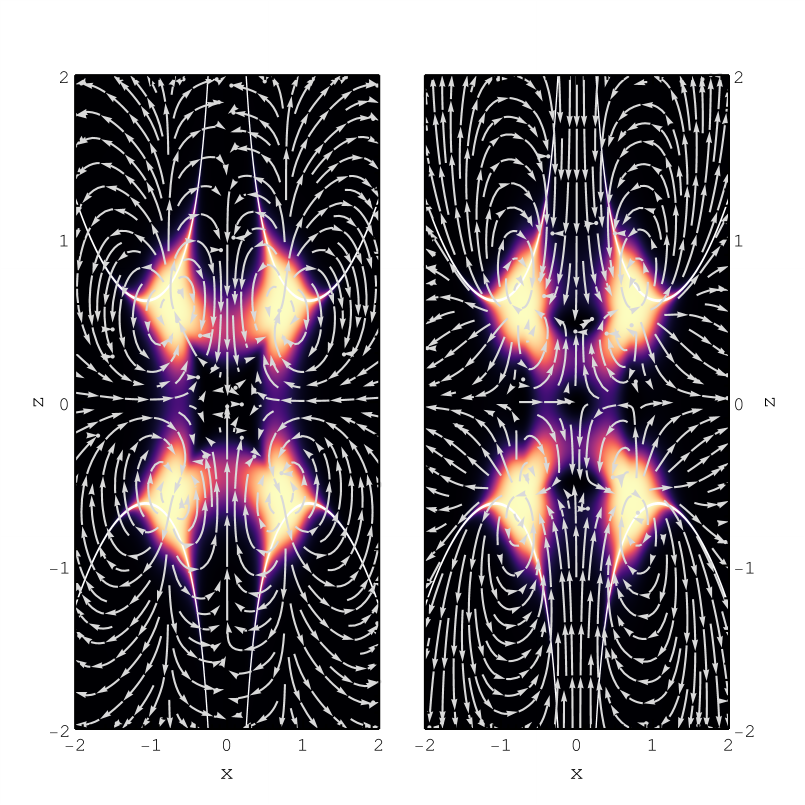}
    \caption{Field lines plotted over the heatmap of (a function of) the norm for the electric field (left panel) and the magnetic field (right panel). The heatmaps show the values of $\operatorname{atan}|\mathbf{E}|$ (left) and  $\operatorname{atan}|\mathbf{B}|$ (right). The lighter the color, the larger the number. Pure white color denotes an infinite vector magnitude.  Plots are for $q_e=-3$, $q_m=-1/2$, and $j=-1/2$.}
    \label{fig:FieldsCase4}
\end{figure}

Concerning the electric and magnetic fields in this electrovac solution, their expressions are quite lengthy, and we prefer to plot them instead. We do so in Fig.~\ref{fig:FieldsCase4}. 
There, we see that the fields decay at infinity and that their magnitude is everywhere finite besides on the ergosurfaces, the latter indeed appearing as poles in the expressions of the norms. We mention here that the graphic inspection is certainly not sufficient, and that the behavior of the electric and magnetic fields in this spacetime deserves further investigation which we will unfortunately not include in this work.

\section{Conclusions}
\label{sec:Conclusions}

This paper focused on obtaining the complete list of stationary and axisymmetric spacetimes, generated from Minkowski spacetime by operating on the associated seed potentials with a composition of Ehlers and Harrison transformations. Adopting the terminology laid down in Sec.~\ref{sec:ErnstForma}, it is clear that the metric functions in the electric and magnetic forms of the WLP metric are completely different functions, though ultimately related via function redefinitions. This implies that the Ernst potentials in the two cases are also going to differ, and thus, any symmetry transformation in potential space is going to provide us also with different target potentials, and ultimately different spacetimes. This is especially clarified in the case of a Minkowski seed, where Ehlers or Harrison transformations of the electric potentials eventually lead us again to seed spacetime (in cylindrical coordinates), modulo rescalings of the noncompact coordinates. On the contrary, an Ehlers or Harrison transformation of the magnetic potentials gives rise to the swirling or electromagnetic universe, respectively. Hence, the form of the WLP metric, electric or magnetic, with which we identify our seed metric, crucially determines the output of the generating process, a deep-seated fact already.

Consequently, taking into account that there are two ways to identify a seed metric and two kinds of transformations we are interested in, we should a priori expect sixteen different spacetimes. However, our seed is Minkowski and thus, this number quickly reduces to ten because a composition of electric transformations does nothing, and $U_{m}\circ U_{e}\sim U_{m}$ where $U=\{\operatorname{E},\operatorname{H}\}$. Recalling that $\operatorname{E}_m\circ\operatorname{E}_m\sim \operatorname{E}_m$, $\operatorname{H}_m\circ\operatorname{E}_m\sim \operatorname{E}_m\circ\operatorname{H}_m$, and that $\operatorname{H}_m\circ\operatorname{H}_m$ is a particular case of $\operatorname{E}_m\circ\operatorname{H}_m$, while also excluding cases where the resulting spacetime is either the swirling or the electromagnetic universe, we should be expecting at most five distinct target spacetimes. Indeed, one of them was extensively studied in Sec.~\ref{sec:EMS}, and the other four in Sec.~\ref{sec:Mix}.

In Sec.~\ref{sec:MS}, we started by reviewing the electromagnetic universe (the extension of the Bonnor--Melvin solution including an additional external electric field) and the swirling solution. These can be obtained from a Minkowski spacetime by operating on its associated magnetic seed potentials with a Harrison and an Ehlers transformation, respectively. Since our first goal was to combine these two, it was reasonable to expect that the resulting spacetime correctly reduces to its building blocks in the appropriate limits, and that it also inherits properties from both. Indeed, applying a magnetic ``Ehlers of a Harrison'' map, we were led to the electromagnetic swirling universe, which met the above expectations. We first probed for surfaces where the metric functions change sign. This revealed the presence of a timelike surface, on which $\partial_t$ is null. Had we been able to single out $\partial_t$ as a unique timelike Killing vector, we would call this \textit{the} ergosurface. However, we pointed out that the usual selection mechanism one uses in Kerr geometry, for example, is not applicable here; there is simply no KV, neither timelike nor normalized, everywhere at infinity. Again, this was not necessarily problematic, since also in the case of Kerr, the definition of $t$ as a ``time'' is meaningful outside the ergosphere; it just happens that here the ergoregions eventually extend to infinity in particular directions. 

We then pondered whether---excluding these regions at infinity where no KV can be timelike---there are other sensible criteria, specific to our case, which can be used to advocate for the existence of a single special timelike Killing vector, to which we should adapt our coordinates. Without being able to rigorously prove that there is no such set of conditions, we nevertheless were not able to find one. The best we could do is to restrict the list of candidates to the one-parameter family $\partial_t+C\partial_\phi$, with $C$ being an arbitrary real parameter. It was then natural to argue that notions such as ergosurfaces, ergoregions, and frame-dragging angular velocity, should be understood with respect to the complete one-parameter family and not just $\partial_t$; the latter is as special as any other member of the former. Moreover, taking the above into consideration, we emphasized that rotation in these spacetimes can only be perceived in a relative manner. The frame-dragging angular velocity $\Omega$, defined with respect to the timelike KV $\partial_t+C\partial_\phi$ in eq.~\eqref{eq:FDAVoneparam}, cannot provide an absolute measure, for there is a free parameter $C$ roaming around. However, we remarked that differences (in the value of $\Omega$ on two different $(\rho,z)$ surfaces) are independent of $C$, and indeed, one may use these differences to (i) prove the existence of rotating regions in the first place, and (ii) to further probe for counterrotating regions, with their exact localization being observer-dependent. Notwithstanding this interesting observation, relevant also in the swirling case (though not addressed so far), we eventually decided to adapt our coordinates to a $C=0$ observer for clarity and simplicity. However, the particular issue definitely requires further investigation, a task we plan to undertake in future work. 

Next, we showed that the EMS spacetime is free of curvature singularities, a Misner string, and conical singularities. We also verified, by checking the components of the Riemann tensor in the suitable orthonormal basis of App.~\ref{app:ON&CNtet}, that tidal forces diminish at infinity in all directions. Actually, the EMS metric is asymptotic to a swirling metric, given that the growth rate of the ratio $\rho/z$ satisfies a certain inequality. However, the complete solution is not, for the gauge field strength does not in general vanish. We applied the method described in Sec.~\ref{sec:Petrov} to conclude that the Petrov type of the solution is D, and then we gave a detailed sequence of coordinate transformations and parameter redefinitions, which ultimately proves that the EMS spacetime belongs to the Kundt family of solutions admitting a shearfree, nonexpanding, and nontwisting null geodesic congruence. The electric and magnetic fields in this spacetime were found to decay far away from the symmetry axis, but, exactly as in the case of the electromagnetic universe, they were uniform in the proximity of the latter, meaning that they do not vanish far away from the equator near the axis. 

Motivated by the well-established relations of the Bonnor--Melvin solution to a planar Reissner--Nordstr\"om spacetime, and of the swirling solution to a planar Taub--NUT spacetime, we managed to successfully demonstrate the relation of our EMS solution to a planar Reissner--Nordstr\"om--NUT spacetime; we did so by performing a double Wick rotation, and by appropriately redefining our coordinates and parameters. This relation, interesting in its own right, proves to be even more intriguing, if one notices that it can be exploited to directly derive the cosmological extension of the solution. The presence of $\Lambda$ renders the field equations into a set of inhomogeneous equations, and the system is no longer integrable. The generating technique no longer applies, for the potential-space symmetries are lost. Thus, the only course of action practically is direct integration, which can be a daunting task. Remarkably, starting from a planar Reissner--Nordstr\"om--NUT spacetime with a cosmological constant, and applying the previous transformations and redefinitions in reverse order, we arrive at a new spacetime, which for $\Lambda=0$ reduces to the EMS spacetime, and which also is a solution to the field equations of the theory. We study this spacetime in some detail, highlighting the presence of a Misner string which can fortunately be exorcised if we properly tune our parameters. Note that the string exists also in the swirling-$\Lambda$ solution of~\cite{Astorino:2022aam}, despite having been overlooked therein. The crucial difference is that in the latter solution, which can be recovered from the EMS-$\Lambda$ solution we presented here in the limit $X\to 0$, there is absolutely no way to ``banish'' it.  

Next, we immersed a Schwarzschild black hole in the EMS background. The presence of a mass source in the seed brought in a singularity and an event horizon also in the target spacetime, their locations not modified. We made it clear that whilst the surface area of the horizon does not change, the surface itself is deformed, with the deformation controlled by the transformation parameters. The structure of the ergoregions was qualitatively the same, while no topological singularities or closed timelike curves were present. The asymptotic behavior of the black hole was that of the background. The solution was found to have Petrov type I almost everywhere besides the axis, where the type was D, and the horizon surface, where we proved that it is algebraically special. We remarked that in order to determine the exact Petrov type on the horizon, a frame attached to a freely falling observer is needed.  

In the second part of this manuscript, we extracted all the spacetimes one can obtain obtain from Minkowski spacetime, by operating on its associated seed potentials with one electric and one magnetic transformation, Ehlers or Harrison. We showed that the order of operations matters. We excluded cases with $U_m\circ U_e$ for reasons previously explained, though we remark that one must consider these when the seed spacetime is other than Minkowski. For example, given a Schwarzschild seed, a combination $\operatorname{U}_m\circ \operatorname{U}_e$ would lead to a completely different spacetime. We registered four novel type I asymptotically nontrivial spacetimes: 
\begin{enumerate}[label=\Roman*.]
    \item A four-parameter family containing the real parameters $E,B$, found in the electromagnetic universe, and two additional real parameters $q_e,q_m$ introduced via the electric Harrison transformation.
    \item A three-parameter family containing $E,B$ and one additional real parameter $c$ entering via the electric Ehlers transformation.
    \item A two-parameter family containing $j$ and one additional parameter $c$ introduced via the electric Ehlers transformation.
    \item A three-parameter family containing $j$, found in the swirling spacetime, and two additional parameters $q_e,q_m$ brought in by the electric Harrison transformation.
\end{enumerate}
As expected, the new spacetimes are complicated modifications of either the electromagnetic universe or the swirling solution. However, these are neither NUTty nor charged, extensions of the latter. To be precise, there is at least no direct evidence to support the association of $c$ with a NUT parameter, or the association of $q_e,q_m$ with monopolic charges, and the ``physical'' meaning of these parameters deserves further scrutiny. 

Starting with the electrovac spacetime I, here the interaction of $q_e,q_m$ with the parameters $E,B$ controlling the magnitude of the external electric and magnetic fields in the electromagnetic universe, remarkably produced a stationary spacetime without ergoregions. We argued that for $q_e^2+q_m^2\leq 4$, the spacetime describes a naked singularity (a ring singularity on the equatorial plane which can be contracted to a point at $\rho=0=z$ when the bound is saturated), otherwise it is regular. We then showcased the presence of a spinning string on the axis which was removable, and once we took care of that, we showed that there are no further topological singularities to be bothered with. Limits of the new geometry to Minkowski spacetime and the electromagnetic universe were checked. By looking at the frame-dragging angular velocity (in coordinates adapted to $\partial_t$), we deduced the presence of counterrotating regions and of surfaces where the it cannot even be defined. These last surfaces were then identified with the so-called chronology horizons which bound the nonchronal regions appearing in this spacetime (regions filled with CTCs). Indeed, we claimed that such regions are expected in the configurations we study if the angular velocity is singular on some surface, and a physical interpretation was given, namely that very close to these surfaces, light cones are completely tipped in the direction of angular motion due to the extreme intensity of rotation building up there. The electromagnetic fields in the target spacetime, finite everywhere, were found to decay in all directions at infinity, a behavior contrasting the one in the electromagnetic universe, where the fields appear uniformly close to the axis. 

Electrovac spacetime II can be thought of as another modification of the electromagnetic universe. We showed that this is yet another stationary background without ergoregions, but with nonchronal regions which exist for arbitrary values of the involved parameters. The electric Ehlers parameter $c$ has nontrivially modified the gauge field of the electromagnetic universe, with the target electric and magnetic fields acquiring an additional component in the $\hat{\rho}$ direction. They again were everywhere nonsingular, and they vanished in all directions at infinity. We remarked that there exists a $(\rho,z)$ surface, on which the fields behave exactly like in the electromagnetic universe. Next, we extracted vacuum spacetime III, practically a modification of the swirling universe. We showed that this stationary spacetime inherited its ergoregion structure from the swirling solution and that it was regular, at least in the sense of regular curvature invariants up to some low order. The absence of topological singularities was verified. As in the preceding backgrounds, this one also featured nonchronal regions, with their geometry however being quite different. In fact, we showed that there can be up to four disconnected regions with CTCs. We proved that there are always two of them, which extend to infinity in certain directions, and that for certain values of the parameters, two additional toroidal regions can appear closer to $z=0$. The fact that the ergoregions partially overlap with the nonchronal regions, made this particular solution even more puzzling. It was shown that there are exactly two rings where chronology horizons and ergosurfaces meet. Finally, electrovac spacetime IV was registered last. For $q_e^2+q_m^2=4$, we found that the new solution, with ergoregions as in the swirling case, describes a naked singularity sitting at $\rho=0=z$, otherwise, it is regular. Topological defects were again absent, but CTCs were present, with their structure being more or less the same as in the case of spacetime III. The expressions of the electromagnetic fields were again too lengthy to explicitly write them down, and thus we plotted them instead. For the corresponding figure, we were able to tell that the fields fall off asymptotically, but that they are not regular everywhere. In particular, we observed that their norm is singular on the ergosurfaces. The specific behavior of the field lines definitely requires a deeper study in order to conclude whether this electromagnetic setup can be eventually of some use. For example, and very roughly speaking, one could say that the plots are somehow reminiscent of two current loops at a given distance, perhaps in the presence of other external electric and magnetic fields, but a convincing explanation of the particular singular surfaces where the fields cannot even be defined definitely eludes us.

We saw that a common factor in all of the new type I solutions in Sec.~\ref{sec:Mix}, was the presence of nonchronal parts. The ``physical'' interpretation given for the emergence of CTCs in the case of spacetime I, also applies to the other cases; their occurrence seems to be, causally enough, related to the fact that the frame-dragging angular velocity (which would also be the velocity of a zero angular momentum observer) blows up, not in some asymptotic region, but actually in finite regions of these spacetimes. As mentioned in the introduction, the problem of CTCs/CNCs requires a much deeper investigation, i.e., whether they are geodesics or, in any case, whether they are actually traversable under sensible conditions. The whole topic is delicate anyway. Protection mechanisms~\cite{SHchrono,NovikoSelf} forbidding causality violation have yet to acquire the form of formal theorems, and it also has been conjectured that causality violating curves exist only in the classical theory (and would disappear in the quantum version). Nevertheless, at the classical level these pathologies exist, and the very fact that they beset honest solutions to the Einstein--Maxwell field equations (or even to pure general relativity), namely that they do not arise due to some artificial distribution of matter or some weird gluing of spacetimes, actually makes them interesting in our opinion, though from a very different perspective. 

The questions opened are perhaps more than the ones answered, and this implies that there is still room and need for further exploration. It would be interesting to see if and how, for example, the consideration of different seed spacetimes would affect fundamental properties of the target spacetimes, obtained via this combination of electric and magnetic transformations. One simple example is the Schwarzschild black hole embedded into spacetime I, which can be thought of as the counterpart of the Reissner--Nordstr\"om black hole immersed in the electromagnetic universe~\cite{Gibbons:2013yq}, in the sense that the latter can be obtained from a Schwarzschild seed via the combination $\operatorname{H}_m\circ \operatorname{H}_e$. In the former case, it is expected that the solution, representing a black hole with a deformed horizon, will not be free of CTCs in the exterior, while it will not feature ergoregions at all. On the other hand, nonchronal regions are absent in~\cite{Gibbons:2001sx}, while ergoregions are present, developing towards asymptotic infinity in some directions. These are two spacetimes with vastly different properties, and with parameters whose physical meaning most probably differs greatly. It is then compelling to deeper understand how and why a simple switch in the order of operations, can bring about so drastically different results, and to shed light on the physical meaning of the parameters in the spacetimes listed here, if any. The existence of a one-parameter family of equally good timelike Killing vectors, instead of a unique one, is also something that requires further research. Moreover, when dealing with purely electric transformations, it is by now understood that any combination of more than two transformations will not yield something new. This does not seem to be the case when mixing transformations, and it is certainly interesting to see if there is a particular number of mixed transformations which gives the same spacetime, regardless of the ordering. Finally, based on the fact that the emergence of CTCs/CNCs proved to be systematic, it is worth investigating---besides scrutinizing the corresponding regions in the spacetimes presented here---whether a set of conditions for the seed spacetime can be mathematically formulated, under which such pathologies can be avoided when applying these transformations.

\acknowledgments
The authors would like to thank Mokhtar Hassa\"{i}ne, Pavel Krtou\v{s}, and Julio Oliva for stimulating discussions. The work of J.B. is supported by FONDECYT Postdoctorado grant 3230596. A.C. is partially supported by FONDECYT Grant 1210500. A.C. and I.K. were supported by Primus grant PRIMUS/23/SCI/005 from Charles University. The work of K.M. is funded by Beca Nacional de Doctorado ANID grant No. 21231943. The work of M.O. is partially funded by Beca ANID de Doctorado grant No. 21222264. K.P. acknowledges financial support provided by the European Regional Development Fund (ERDF) through the Center of Excellence TK133 ``The Dark Side of the Universe'' and PRG356 ``Gauge gravity: unification, extensions and phenomenology''. K.P. also acknowledges participation in the COST Association Action CA18108 ``Quantum Gravity Phenomenology in the Multimessenger Approach (QG-MM)''.

\appendix
\section{Orthonormal frames and complex null tetrads\label{app:ON&CNtet}}

Given the WLP metric~\eqref{eq:eleWLP}, we choose an orthonormal coframe $\{\vartheta^{\mathbf{a}}\}$ with 
\begin{equation}
    \vartheta^{\mathbf{0}}=\sqrt{f}(dt-\omega\,d\phi),\quad \vartheta^{\mathbf{1}}=\mathrm{e}^{\gamma}f^{-1/2}d\rho,\quad \vartheta^{\mathbf{2}}=\mathrm{e}^{\gamma}f^{-1/2}dz,\quad \vartheta^{\mathbf{3}}=\rho f^{-1/2}d\phi,
\end{equation}
dual to the basis $\{e_{\mathbf{a}}\}$ with 
\begin{equation}
    e_{\mathbf{0}}=f^{-1/2}\partial_t,\quad e_{\mathbf{1}}=\sqrt{f}\mathrm{e}^{-\gamma}\partial_\rho,\quad e_{\mathbf{2}}=\sqrt{f}\mathrm{e}^{-\gamma}\partial_z,\quad e_{\mathbf{3}}=\sqrt{f}\rho^{-1}(\partial_\phi+\omega\partial_t).
\end{equation}
Any tensor $T^{\mu\ldots\nu}{}_{\lambda\ldots\rho}$ can be written with frame indices as
\begin{equation}
T^{\mathbf{a}\ldots\mathbf{b}}{}_{\mathbf{c}\ldots\mathbf{d}}=T^{\mu\ldots\nu}{}_{\lambda\ldots\rho}\vartheta_\mu^{\mathbf{a}}\ldots \vartheta_\nu^{\mathbf{b}}e^\lambda_{\mathbf{c}}\ldots e^\rho_{\mathbf{d}}.
\end{equation}

We can construct an initial complex null tetrad $\{\mathfrak{e}_a\}=\{k,l,m,\overline{m}\}$ by taking combinations of the frame ``legs'' $e_{\mathbf{a}}$. In particular,  
\begin{equation}
    \begin{split}
        \sqrt{2}k&=e_{\mathbf{0}}-e_{\mathbf{3}},\\
        \sqrt{2}l&=e_{\mathbf{0}}+e_{\mathbf{3}},\\
        \sqrt{2}m&=e_{\mathbf{2}}+ie_{\mathbf{1}}.
    \end{split}
\end{equation}
At this stage, we perform a local Lorentz transformation of the null basis with a matrix 
\begin{equation}
    \Lambda^{-1}(\rho,z) = \operatorname{diag}(1/k^0,k^0,1,1),
\end{equation}
to arrive at
\begin{equation}
    \mathfrak{e}' = \Lambda^{-1}\mathfrak{e} = \{ k/k^0,k^0 l,m,\overline{m}\}=:\{k',l',m,\overline{m}\}.
\end{equation}
The new basis has the nice property that $\Psi'_4\neq 0$ and
\begin{equation}
    \Psi'_1\equiv C_{\lambda\rho\mu\nu}k'^\lambda l'^\rho k'^\mu m^\nu=0=C_{\lambda\rho\mu\nu}k'^\lambda l'^\rho \overline{m}^\mu l'^\nu\equiv\Psi_3,
\end{equation}
in the general case, which simplifies the Petrov classification a lot. 

Since we display the black hole spacetimes in spherical-like coordinates $\{t,r,x,\phi\}$ for convenience, we also provide our choice of orthonormal coframe in the latter coordinate system:
\begin{equation}
    \begin{split}
        \vartheta^{\mathbf{0}}&=\sqrt{f}(dt-\omega\,d\phi),\\
        \vartheta^{\mathbf{1}}&=\mathrm{e}^{\gamma}\sqrt{(\partial \rho/\partial r)^2+(\partial z/\partial r)^2}f^{-1/2} dr,\\
        \vartheta^{\mathbf{2}}&=\mathrm{e}^{\gamma}\sqrt{(\partial \rho/\partial r)^2+(\partial z/\partial r)^2}f^{-1/2}(\partial z/\partial r)^{-1}(\partial \rho/\partial x) dx,\\
        \vartheta^{\mathbf{3}}&=\rho f^{-1/2} d\phi.
    \end{split}
\end{equation}
Here, $f,\omega,\gamma,\rho,z$ are functions of $r,x$, and the tetrad $e$ is given by the inverse transpose of $\vartheta$, namely $e=(\vartheta^T)^{-1}$. The desired CNT can then be constructed using the previously demonstrated recipe. 

For the metric~\eqref{eq:mWLP}, we choose our orthonormal coframe as 
\begin{equation}
    \vartheta^{\mathbf{0}}=\rho f^{-1/2}dt,\quad \vartheta^{\mathbf{1}}=\mathrm{e}^{\gamma}f^{-1/2}d\rho,\quad \vartheta^{\mathbf{2}}=\mathrm{e}^{\gamma}f^{-1/2}dz,\quad \vartheta^{\mathbf{3}}=\sqrt{f}(d\phi-\omega\,dt).
\end{equation}
Its dual, the orthonormal basis $\{e_{\mathbf{a}}\}$, is comprised of 
\begin{equation}\label{eq:ONTmag}
   e_{\mathbf{0}}=\sqrt{f}\rho^{-1}(\partial_t+\omega\partial_\phi),\quad e_{\mathbf{1}}=\sqrt{f}\mathrm{e}^{-\gamma}\partial_\rho,\quad e_{\mathbf{2}}=\sqrt{f}\mathrm{e}^{-\gamma}\partial_z,\quad  e_{\mathbf{3}}=f^{-1/2}\partial_\phi.
\end{equation}
To construct the desired CNT, the one for which $\Psi_1=0=\Psi_3$ and $\Psi_4\neq 0$ in the general case, we follow the previous prescription, arriving at 
\begin{equation}
\begin{split}
    k &= \partial_t+(\omega-\rho f^{-1})\partial_\phi,\\
    2 l &= \rho^{-2}[f\,\partial_t +(\rho+f\omega)\partial_\phi],\\
    \sqrt{2} m &= \sqrt{f}\mathrm{e}^{-\gamma}(\partial_z+i\partial_\rho).
    \end{split}
\end{equation}
Finally, when using spherical-like coordinates, we choose the cobasis
\begin{equation}
    \begin{split}
        \vartheta^{\mathbf{0}}&=\rho f^{-1/2} dt,\\
        \vartheta^{\mathbf{1}}&=\mathrm{e}^{\gamma}\sqrt{(\partial \rho/\partial r)^2+(\partial z/\partial r)^2}f^{-1/2} dr,\\
        \vartheta^{\mathbf{2}}&=\mathrm{e}^{\gamma}\sqrt{(\partial \rho/\partial r)^2+(\partial z/\partial r)^2}f^{-1/2}(\partial z/\partial r)^{-1}(\partial \rho/\partial x) dx,\\
        \vartheta^{\mathbf{3}}&=\sqrt{f}(d\phi-\omega\,dt),
    \end{split}
\end{equation}
where the involved functions are functions of $r,x$. Again, the CNT is constructed exactly in the previous fashion. 

\section{Electric and magnetic fields\label{app:EMfields}}
In this appendix section, we would like to say a few words about the method used to extract the electric and magnetic fields. Instead of using gradient and curl operators, we simply find the components of the fields from the electromagnetic tensor $F_{\mathbf{ab}}$. Since bold latin indices are raised/lowered with the Minkowski metric $\eta$, it follows that the electromagnetic tensor has the form 
\begin{equation}
    F = -E_{\mathbf{i}}\vartheta^{\mathbf{0}}\wedge\vartheta^{\mathbf{i}}+\frac{1}{2}\epsilon_{\mathbf{ijk}}B^{\mathbf{i}}\vartheta^{\mathbf{j}}\wedge\vartheta^{\mathbf{k}},
\end{equation}
where $\mathbf{i,j,k,\ldots}=\mathbf{1,2,3}$, and $\epsilon_{\mathbf{ijk}}=\epsilon_{\mathbf{0abc}}$ with $\epsilon_{\mathbf{abcd}}$ being the four-dimensional Levi-Civita tensor. Note that $\eta_{\mathbf{ij}}=\delta_{\mathbf{ij}}$, and thus $E_{\mathbf{i}}=E^{\mathbf{i}}$; ditto for the components of the magnetic field. Due to the form of the gauge field, the electric and magnetic fields do not admit an azimuthal component, and they can be written in terms of the unit basis $\{\hat{\rho},\hat{z},\hat{\phi}\}$ as
\begin{equation}
    \mathbf{E}=E^{\mathbf{i}}e_{\mathbf{i}} = E^{\mathbf{1}} e_{\mathbf{1}}^\rho\sqrt{g_{\rho\rho}}\,\hat{\rho}+E^{\mathbf{2}} e_{\mathbf{2}}^z\sqrt{g_{zz}}\,\hat{z}=E^{\mathbf{1}}\hat{\rho}+E^{\mathbf{2}}\hat{z}.
\end{equation}
The basis expansion of $\mathbf{B}$ can of course be obtained by replacing $E$ with $B$ in the above. These formulas are valid for both, electric and magnetic, forms of the metric. 

For the electric case, we find that 
\begin{equation}
    \mathbf{E}=\mathrm{e}^{-\gamma} \bs{\nabla} A_t,\quad \mathbf{B}=\mathrm{e}^{-\gamma}\bs{\nabla}\widetilde{A}_\phi,
\end{equation}
where the gradient is understood as the flat one in cylindrical coordinates, and the twist potential $\widetilde{A}_\phi$ is defined via eq.~\eqref{eq:eqATphiE}. The Riemann--Silberstein vector is thus
\begin{equation}
    \mathbf{X}=\mathbf{E}+i\mathbf{B}=\mathrm{e}^{-\gamma}\bs{\nabla}\Phi,
\end{equation}
with $\Phi$ given in~\eqref{eq:ElePots}. For the magnetic case, we have 
\begin{equation}
    \mathbf{E}=\mathrm{e}^{-\gamma}\hat{\phi}\times \bs{\nabla} \widetilde{A}_t,\quad \mathbf{B}=-\mathrm{e}^{-\gamma}\hat{\phi}\times \bs{\nabla} A_\phi.
\end{equation}
Thus, the Riemann--Silberstein vector is now 
\begin{equation}
    \mathbf{X}=-i\mathrm{e}^{-\gamma}\hat{\phi}\times \bs{\nabla}\Phi,
\end{equation}
where $\Phi$ is given in eq.~\eqref{eq:MagPots}. As a cross-check, one can verify that in both cases 
\begin{equation}
    \frac{1}{2}F_{\mu\nu}F^{\mu\nu}=-\Im(i\mathbf{X}\cdot\mathbf{X})=\mathbf{B}\cdot \mathbf{B}-\mathbf{E}\cdot\mathbf{E}
\end{equation}
\bibliographystyle{JHEP}
\bibliography{biblio}% Produces the bibliography via BibTeX.

\providecommand{\href}[2]{#2}\begingroup\raggedright\begin{thebibliography}{10}

\bibitem{Stephani:2003tm}
H.~Stephani, D.~Kramer, M.A.H.~MacCallum, C.~Hoenselaers and E.~Herlt, \emph{{Exact solutions of Einstein's field equations}}, Cambridge Monographs on Mathematical Physics, Cambridge Univ. Press, Cambridge (2003), \href{https://doi.org/10.1017/CBO9780511535185}{10.1017/CBO9780511535185}.

\bibitem{Bekenstein:1972tm}
J.D.~Bekenstein, \emph{{Black holes and the second law}}, \href{https://doi.org/10.1007/BF02757029}{\emph{Lett. Nuovo Cim.} {\bfseries 4} (1972) 737}.

\bibitem{Bekenstein:1973ur}
J.D.~Bekenstein, \emph{{Black holes and entropy}}, \href{https://doi.org/10.1103/PhysRevD.7.2333}{\emph{Phys. Rev. D} {\bfseries 7} (1973) 2333}.

\bibitem{Hawking:1974rv}
S.W.~Hawking, \emph{{Black hole explosions}}, \href{https://doi.org/10.1038/248030a0}{\emph{Nature} {\bfseries 248} (1974) 30}.

\bibitem{Hawking:1975vcx}
S.W.~Hawking, \emph{{Particle Creation by Black Holes}}, \href{https://doi.org/10.1007/BF02345020}{\emph{Commun. Math. Phys.} {\bfseries 43} (1975) 199}.

\bibitem{Giddings:1995gd}
S.B.~Giddings, \emph{{The Black hole information paradox}},  in \emph{{PASCOS / HOPKINS 1995 (Joint Meeting of the International Symposium on Particles, Strings and Cosmology and the 19th Johns Hopkins Workshop on Current Problems in Particle Theory)}}, pp.~415--428, 8, 1995 [\href{https://arxiv.org/abs/hep-th/9508151}{{\ttfamily hep-th/9508151}}].

\bibitem{Susskind:1994vu}
L.~Susskind, \emph{{The World as a hologram}}, \href{https://doi.org/10.1063/1.531249}{\emph{J. Math. Phys.} {\bfseries 36} (1995) 6377} [\href{https://arxiv.org/abs/hep-th/9409089}{{\ttfamily hep-th/9409089}}].

\bibitem{Ehlers:1957zz}
J.~Ehlers, \emph{{Konstruktionen und Charakterisierung von Losungen der Einsteinschen Gravitationsfeldgleichungen}},  thesis, 1957.

\bibitem{Ehlers:1959aug}
J.~Ehlers, \emph{{Transformations of static exterior solutions of Einstein's gravitational field equations into different solutions by means of conformal mapping}}, {\emph{Colloq. Int. CNRS} {\bfseries 91} (1962) 275}.

\bibitem{harrison1968new}
B.K.~Harrison, \emph{New solutions of the einstein-maxwell equations from old}, {\emph{Journal of Mathematical Physics} {\bfseries 9} (1968) 1744}.

\bibitem{Ernst:1967wx}
F.J.~Ernst, \emph{{New formulation of the axially symmetric gravitational field problem}}, \href{https://doi.org/10.1103/PhysRev.167.1175}{\emph{Phys. Rev.} {\bfseries 167} (1968) 1175}.

\bibitem{Ernst:1967by}
F.J.~Ernst, \emph{{New Formulation of the Axially Symmetric Gravitational Field Problem. II}}, \href{https://doi.org/10.1103/PhysRev.168.1415}{\emph{Phys. Rev.} {\bfseries 168} (1968) 1415}.

\bibitem{NeugeKramer69}
D.~{Kramer} and G.~{Neugebauer}, \emph{{Eine exakte station{\"a}re L{\"o}sung der EINSTEIN-MAXWELL-Gleichungen}}, \href{https://doi.org/10.1002/andp.19694790107}{\emph{Annalen der Physik} {\bfseries 479} (1969) 59}.

\bibitem{Kinn1}
W.~{Kinnersley}, \emph{{Generation of stationary Einstein-Maxwell fields}}, \href{https://doi.org/10.1063/1.1666373}{\emph{Journal of Mathematical Physics} {\bfseries 14} (1973) 651}.

\bibitem{Geroch}
R.~Geroch, \emph{Method for generating solutions of einstein's equations.}, \href{https://doi.org/10.1063/1.1665681}{\emph{J. Math. Phys. (N. Y.) 12: No. 6, 918-24(Jun 1971).} (1971) }.

\bibitem{Israel:1972vx}
W.~Israel and G.A.~Wilson, \emph{{A class of stationary electromagnetic vacuum fields}}, \href{https://doi.org/10.1063/1.1666066}{\emph{J. Math. Phys.} {\bfseries 13} (1972) 865}.

\bibitem{Barrientos:2023tqb}
J.~Barrientos and A.~Cisterna, \emph{{Ehlers transformations as a tool for constructing accelerating NUT black holes}}, \href{https://doi.org/10.1103/PhysRevD.108.024059}{\emph{Phys. Rev. D} {\bfseries 108} (2023) 024059} [\href{https://arxiv.org/abs/2305.03765}{{\ttfamily 2305.03765}}].

\bibitem{Astorino:2023elf}
M.~Astorino and G.~Boldi, \emph{{Plebanski-Demianski goes NUTs (to remove the Misner string)}}, \href{https://doi.org/10.1007/JHEP08(2023)085}{\emph{JHEP} {\bfseries 08} (2023) 085} [\href{https://arxiv.org/abs/2305.03744}{{\ttfamily 2305.03744}}].

\bibitem{Astorino:2023ifg}
M.~Astorino, \emph{{Accelerating and charged type I black holes}}, \href{https://doi.org/10.1103/PhysRevD.108.124025}{\emph{Phys. Rev. D} {\bfseries 108} (2023) 124025} [\href{https://arxiv.org/abs/2307.10534}{{\ttfamily 2307.10534}}].

\bibitem{Barrientos:2023dlf}
J.~Barrientos, A.~Cisterna and K.~Pallikaris, \emph{{Pleban\'ski-Demia\'nski \`a la Ehlers-Harrison: Exact Rotating and Accelerating Type I Black Holes}},  \href{https://arxiv.org/abs/2309.13656}{{\ttfamily 2309.13656}}.

\bibitem{newman1961new}
E.~Newman and L.~Tamburino, \emph{New approach to einstein's empty space field equations}, {\emph{Journal of Mathematical Physics} {\bfseries 2} (1961) 667}.

\bibitem{robinson1962some}
I.~Robinson and A.~Trautman, \emph{Some spherical gravitational waves in general relativity}, {\emph{Proceedings of the Royal Society of London. Series A. Mathematical and Physical Sciences} {\bfseries 265} (1962) 463}.

\bibitem{witten1962gravitation}
L.~Witten, \emph{Gravitation: An Introduction to Current Research}, Wiley (1962).

\bibitem{BachWeyl}
R.~Bach and W.~H, \emph{{Neuelösungen der einsteinschen gravitationsgleichungen}}, {\emph{Math.Z} {\bfseries 13} (1922) 134}.

\bibitem{Bonnor_1954}
W.B.~Bonnor, \emph{Static magnetic fields in general relativity}, \href{https://doi.org/10.1088/0370-1298/67/3/305}{\emph{Proceedings of the Physical Society. Section A} {\bfseries 67} (1954) 225}.

\bibitem{Melvin1966}
M.A.~Melvin and J.S.~Wallingford, \emph{Orbits in a magnetic universe}, \href{https://doi.org/10.1063/1.1704937}{\emph{Journal of Mathematical Physics} {\bfseries 7} (1966) 333}.

\bibitem{Astorino:2022aam}
M.~Astorino, R.~Martelli and A.~Vigan\`o, \emph{{Black holes in a swirling universe}}, \href{https://doi.org/10.1103/PhysRevD.106.064014}{\emph{Phys. Rev. D} {\bfseries 106} (2022) 064014} [\href{https://arxiv.org/abs/2205.13548}{{\ttfamily 2205.13548}}].

\bibitem{Gibbons:2013yq}
G.W.~Gibbons, A.H.~Mujtaba and C.N.~Pope, \emph{{Ergoregions in Magnetised Black Hole Spacetimes}}, \href{https://doi.org/10.1088/0264-9381/30/12/125008}{\emph{Class. Quant. Grav.} {\bfseries 30} (2013) 125008} [\href{https://arxiv.org/abs/1301.3927}{{\ttfamily 1301.3927}}].

\bibitem{Vigano:2022hrg}
A.~Vigan\`o, \emph{{Black Holes and Solution Generating Techniques}}, Ph.D. thesis, Milan U., 2022.
\newblock \href{https://arxiv.org/abs/2211.00436}{{\ttfamily 2211.00436}}.

\bibitem{Petrov2000TheCO}
A.Z.~Petrov, \emph{The classification of spaces defining gravitational fields}, {\emph{General Relativity and Gravitation} {\bfseries 32} (2000) 1665}.

\bibitem{dInverno}
R.A.~d'Inverno and R.A.~Russell-Clark, \emph{Classification of the harrison metrics.}, {\emph{Journal of Mathematical Physics} {\bfseries 12} (1971) 1258}.

\bibitem{griffiths_podolsky_2009}
J.B.~Griffiths and J.~Podolský, \emph{Exact Space-Times in Einstein's General Relativity}, Cambridge Monographs on Mathematical Physics, Cambridge University Press (2009), \href{https://doi.org/10.1017/CBO9780511635397}{10.1017/CBO9780511635397}.

\bibitem{Taub:1950ez}
A.H.~Taub, \emph{{Empty space-times admitting a three parameter group of motions}}, \href{https://doi.org/10.2307/1969567}{\emph{Annals Math.} {\bfseries 53} (1951) 472}.

\bibitem{Kinnersley1969TYPEDV}
W.M.~Kinnersley, \emph{Type d vacuum metrics.}, {\emph{Journal of Mathematical Physics} {\bfseries 10} (1969) 1195}.

\bibitem{PateraWinternitz}
J.~{Patera} and P.~{Winternitz}, \emph{{Subalgebras of real three- and four-dimensional Lie algebras}}, \href{https://doi.org/10.1063/1.523441}{\emph{Journal of Mathematical Physics} {\bfseries 18} (1977) 1449}.

\bibitem{Capobianco:2023kse}
R.~Capobianco, B.~Hartmann and J.~Kunz, \emph{{Geodesic Motion in a Swirling Universe: The complete set of solutions}},  \href{https://arxiv.org/abs/2312.17347}{{\ttfamily 2312.17347}}.

\bibitem{Illy:2023iau}
M.~Illy, \emph{{Accelerated Reissner-Nordstrom black hole in a swirling, magnetic universe}},  other thesis, 12, 2023, [\href{https://arxiv.org/abs/2312.14995}{{\ttfamily 2312.14995}}].

\bibitem{Frolov:1998wf}
V.P.~Frolov and I.D.~Novikov, eds., \emph{{Black hole physics: Basic concepts and new developments}} (1998), \href{https://doi.org/10.1007/978-94-011-5139-9}{10.1007/978-94-011-5139-9}.

\bibitem{Kundt1961}
W.~{Kundt}, \emph{{The plane-fronted gravitational waves}}, \href{https://doi.org/10.1007/BF01328918}{\emph{Zeitschrift fur Physik} {\bfseries 163} (1961) 77}.

\bibitem{Kundt1962ExactSO}
W.~Kundt, \emph{Exact solutions of the field equations: twist-free pure radiation fields}, {\emph{Proceedings of the Royal Society of London. Series A. Mathematical and Physical Sciences} {\bfseries 270} (1962) 328 }.

\bibitem{Gibbons:2001sx}
G.W.~Gibbons and C.A.R.~Herdeiro, \emph{{The Melvin universe in Born-Infeld theory and other theories of nonlinear electrodynamics}}, \href{https://doi.org/10.1088/0264-9381/18/9/305}{\emph{Class. Quant. Grav.} {\bfseries 18} (2001) 1677} [\href{https://arxiv.org/abs/hep-th/0101229}{{\ttfamily hep-th/0101229}}].

\bibitem{Charmousis:2006fx}
C.~Charmousis, D.~Langlois, D.A.~Steer and R.~Zegers, \emph{{Rotating spacetimes with a cosmological constant}}, \href{https://doi.org/10.1088/1126-6708/2007/02/064}{\emph{JHEP} {\bfseries 02} (2007) 064} [\href{https://arxiv.org/abs/gr-qc/0610091}{{\ttfamily gr-qc/0610091}}].

\bibitem{Zofka:2019yfa}
M.~\v{Z}ofka, \emph{{Bonnor-Melvin universe with a cosmological constant}}, \href{https://doi.org/10.1103/PhysRevD.99.044058}{\emph{Phys. Rev. D} {\bfseries 99} (2019) 044058} [\href{https://arxiv.org/abs/1903.08563}{{\ttfamily 1903.08563}}].

\bibitem{Astorino:2012zm}
M.~Astorino, \emph{{Charging axisymmetric space-times with cosmological constant}}, \href{https://doi.org/10.1007/JHEP06(2012)086}{\emph{JHEP} {\bfseries 06} (2012) 086} [\href{https://arxiv.org/abs/1205.6998}{{\ttfamily 1205.6998}}].

\bibitem{Pravda:2005uv}
V.~Pravda and O.B.~Zaslavskii, \emph{{Curvature tensors on distorted Killing horizons and their algebraic classification}}, \href{https://doi.org/10.1088/0264-9381/22/23/009}{\emph{Class. Quant. Grav.} {\bfseries 22} (2005) 5053} [\href{https://arxiv.org/abs/gr-qc/0510095}{{\ttfamily gr-qc/0510095}}].

\bibitem{ErnstSchwMel}
F.J.~{Ernst}, \emph{{Black holes in a magnetic universe}}, \href{https://doi.org/10.1063/1.522781}{\emph{Journal of Mathematical Physics} {\bfseries 17} (1976) 54}.

\bibitem{SHchrono}
S.W.~Hawking, \emph{Chronology protection conjecture}, \href{https://doi.org/10.1103/PhysRevD.46.603}{\emph{Phys. Rev. D} {\bfseries 46} (1992) 603}.

\bibitem{vanStockum:1937zz}
W.J.~van Stockum, \emph{{The gravitational feild of a distribution of particles rotating about an axis of symmetry}}, {\emph{Proc. Roy. Soc. Edinburgh} {\bfseries 57} (1937) 135}.

\bibitem{BonnorDust1}
W.B.~{Bonnor}, \emph{{The rigidly rotating relativistic dust cylinder}}, \href{https://doi.org/10.1088/0305-4470/13/6/033}{\emph{Journal of Physics A Mathematical General} {\bfseries 13} (1980) 2121}.

\bibitem{NovikoSelf}
J.~{Friedman}, M.S.~{Morris}, I.D.~{Novikov}, F.~{Echeverria}, G.~{Klinkhammer}, K.S.~{Thorne} et~al., \emph{{Cauchy problem in spacetimes with closed timelike curves}}, \href{https://doi.org/10.1103/PhysRevD.42.1915}{\emph{Phys. Rev. D} {\bfseries 42} (1990) 1915}.

\end{thebibliography}\endgroup
\end{document}